# Using satellite imagery to map rural marketplaces and monitor their activity at high frequency

Tillmann von Carnap[1,2]*, Reza M. Asiyabi[3,4], Paul Dingus[2], Anna Tompsett[5,6]

**Affiliations:**

[1] Department of Economics, University of Oslo; Oslo, 0851, Norway
[2] Center on Food Security and the Environment, Stanford University; Stanford, 94305, United States of America.

[3] Mistra Center for Sustainable Markets, Stockholm School of Economics; Stockholm; 11350; Sweden

[4] School of GeoScience, University of Edinburgh; Edinburgh; United Kingdom; EH8 9XP; United Kingdom

[5] Beijer Institute of Ecological Economics, The Royal Swedish Academy of Sciences; Stockholm; 10405; Sweden

[6] Institute for International Economic Studies, Stockholm University; Stockholm; 10691; Sweden

*Corresponding author. tillmanv@econ.uio.no

**Author contributions:**

Conceptualization: TC

Methodology: TC, RA, PD, AT

Investigation: TC, PD, AT

Funding acquisition: TC, AT

Writing – original draft: TC

Writing – review & editing: TC, RA, PD, AT

**Competing interests:** Authors declare that they have no competing interests.

**Classification:** Social Sciences - Sustainability Science

**This PDF file includes:**

Abstract

Main Text and Figures

Funding acknowledgement

References

Supplementary Materials

**Abstract:** In many rural areas of low- and middle-income countries, weekly gatherings of buyers and sellers are the most tangible manifestation of the market economy. Knowing these markets' whereabouts and activity over time could provide insights in otherwise data-scarce environments, helping researchers and policymakers to better understand poor rural economies. But these markets are by nature informal and scattered widely across often-remote regions. As a result, data on this fundamental institution are sparse and inconsistent. We develop, test, and apply a method to fill this gap, leveraging market activity's unique temporal and visual signature in satellite imagery. Using secondary data from Kenya, Malawi, and Mozambique, we first confirm that we detect markets with high sensitivity and specificity. We then derive a map of 1,776 markets in Ethiopia and track their activity at up-to-weekly frequency between 2017 and 2024. Measured market activity exhibits seasonal patterns following local agricultural calendars and responds to weather and conflict shocks. Our approach is applicable wherever satellites can regularly acquire images of rural periodic markets and requires no ground data. Once markets are mapped, our approach can be fully automated to produce an up-to-weekly measure of economic conditions in areas where such data is otherwise generally not available.

## Introduction

Across much of the world, buyers and sellers of goods and services gather to trade in specific locations on specific days. Such periodic markets simplify the identification of trading partners where populations live dispersed, allow for the bulking of goods where individual transport is uneconomical, and enable face-to-face transactions where weak legal systems complicate contracting[1]. The conditions making periodic markets an efficient mode of exchange applied from the dawn of trading systems until the industrial era, and they still do in rural areas of many developing economies today[2–4]. Consequently, periodic markets have been documented across pre-historic societies from the Roman Empire to pre-colonial West Africa to Mughal-era India[5–7], and they remain widespread today.

The same conditions that make it efficient to trade at periodic marketplaces also make data about these marketplaces extremely scarce. Because they are rural and informal, consistent maps are typically not available, nor is market attendance recorded over time. This prevents researchers from gaining a deeper understanding of how these markets evolve and respond to external factors that potentially trigger or impede economic growth. Beyond a better understanding of markets *per se*, changes in activity could also be informative about changes in economic conditions more broadly. Where a large share of local trade occurs at such marketplaces, busier market days likely reflect more income spent, more goods traded, or both.

In this paper, we develop, test and apply a method that addresses these gaps by using high-frequency satellite imagery to map periodic marketplaces and track their activity over time. Our approach requires only satellite imagery and applying it does not depend on the availability of any prior information about market locations. Instead, we rely on markets' highly distinctive visual and temporal signature in time series of images.

To evaluate how successfully we detect markets, we would ideally use up-to-date, internationally comprehensive market maps. Similarly, we would ideally evaluate our activity measure using high-frequency local measures of economic activity. Such data largely do not exist, however, and their absence motivates our work. We make progress by validating our method's ability to detect markets with secondary data from Kenya, Malawi, and Mozambique. We then apply the method to derive a novel map of 1,776 marketplaces across all of Ethiopia outside its capital, Addis Ababa, and an activity panel for these markets between 2017 and 2024. No method could detect *every* market, given that the term can informally describe anything from a handful of individual stalls selling small quantities of goods to large, urban wholesale operations. We detect a particular type of market that is widespread, especially in rural areas: occurring periodically, during daylight hours, and at least partially in the open air. We then show that our measure of market activity correlates within and across seasons with local rainfall, a known determinant of rural economic activity[8]: on average, growing-season rainfall decreasing by one standard deviation is associated with market activity in the subsequent harvest season decreasing by 4.2 index points, corresponding to 24% of the seasonal difference between average market activity and lean season market activity. Lastly, we illustrate how the method can be applied for monitoring during challenging circumstances with an application to recent violent conflict in Ethiopia.

Our approach has several potential applications. Better information about markets' locations could help policymakers design more effective rural development policies, such as improvements to rural transport systems. High-frequency information about market activity could also help generate timely feedback about the success or otherwise of different interventions designed to promote market integration, a stated goal of many development policies. High on-the-ground data collection costs can otherwise make it difficult to evaluate which programs work or to learn how to improve program design. Finally, our data can potentially be integrated into early-warning systems in contexts where on-the-ground information on economic conditions is too costly, or unsafe, to obtain.

More broadly, tracking rural marketplaces with satellite imagery presents unique advantages for monitoring in remote economies, expanding the toolkit available to practitioners[9]. Other sources of information, such as enterprise or household surveys, are infrequently collected, spatially sparse, and costly to scale. In contrast, modern satellite imagery combines high spatial resolution with dense geographical and temporal coverage[10–12]. Absent cloud cover, satellites capture near-daily images of every outdoor marketplace worldwide. Satellites also collect data continuously and consistently regardless of local conditions, ensuring comparability across space and time. Finally, and unlike survey data, satellite images are available in near real-time, and the processing of imagery needed to track activity at a marketplace can be fully automated. This provides an unprecedented ability to monitor rapidly evolving situations, including in conflict-affected settings like the Ethiopian example we study.

Existing applications of satellite data to learn about economic conditions in low-income countries implicitly or explicitly infer conditions from large assets that are visible from space, such as buildings or infrastructure[13,14]. These assets evolve slowly, however, meaning that the derived indicators can be slow to pick up even substantial changes in economic conditions. For example, light emissions at night – a commonly used proxy for economic development – cannot reliably pick up short-term changes[16] nor discern variation in living standards among the poorest[13,15]. Indeed, 63% of marketplaces in our map of Ethiopia have no discernible local light emissions at all. Consequently, existing approaches based on satellite imagery are primarily informative for comparisons across space or over long periods. In contrast, marketplace activity can visibly change every week, providing the spatial and temporal granularity needed to measure short-term responses to economic shocks or policies, or to help target assistance[17].

In summary, this paper demonstrates how to detect marketplaces and track their activity using globally available near-daily satellite imagery. Using widely accessible data, our method can produce temporally and geographically fine-grained measures of a central feature of rural economic activity.

### Detecting and tracking marketplaces

The basic intuition for our approach is that marketplaces appear differently on market days and non-market days, as shown in very-high-resolution (VHR; 30cm) images for a location in Mozambique (Fig. 1A). At this resolution, colourful market structures and attending crowds are clearly visible on a Sunday, the local market day, and conspicuously absent on a Wednesday.

Acquiring such VHR imagery at scale is prohibitively expensive, however, and available images only infrequently capture market days. Instead, we use lower resolution (3.1m) PlanetScope RGB imagery that is globally available with a median average revisit interval of 30 hours[18]. In an individual Sunday image at this lower resolution (Fig. 1B), market activity at the same location is barely identifiable, with the location only appearing slightly darker than in a corresponding Wednesday image. We can nonetheless detect a temporal signature in a time series of images, because appearance differences that are associated with market activity occur *regularly*, on market days. In contrast, other appearance differences, such as those arising from the ploughing of fields or wetting of soils, occur erratically or at other frequencies.

To isolate the periodic appearance differences that characterise market activity from other changes, we would ideally use imagery from the same location on a non-market day as a reference. As we do not know market days beforehand, we instead build on the intuition that, for markets that occur fewer than half the days in a week, a median composite of a sufficiently long time series of spatially aligned images (i.e., the median value of each pixel across the series) will resemble a non-market day. We therefore construct for each image a reference composite using imagery from surrounding dates. Subtracting this composite yields a difference image with areas highlighted that appear unusual for the given period, reducing in particular the influence of seasonal vegetation changes (heatmaps in Fig. 1C). Averaging these difference images by day-of-week yields a representation where pixels with periodic changes on a given day-of-week exhibit higher values than those whose appearance does not vary at this frequency (yellow and red areas in upper row of Fig. 1D).

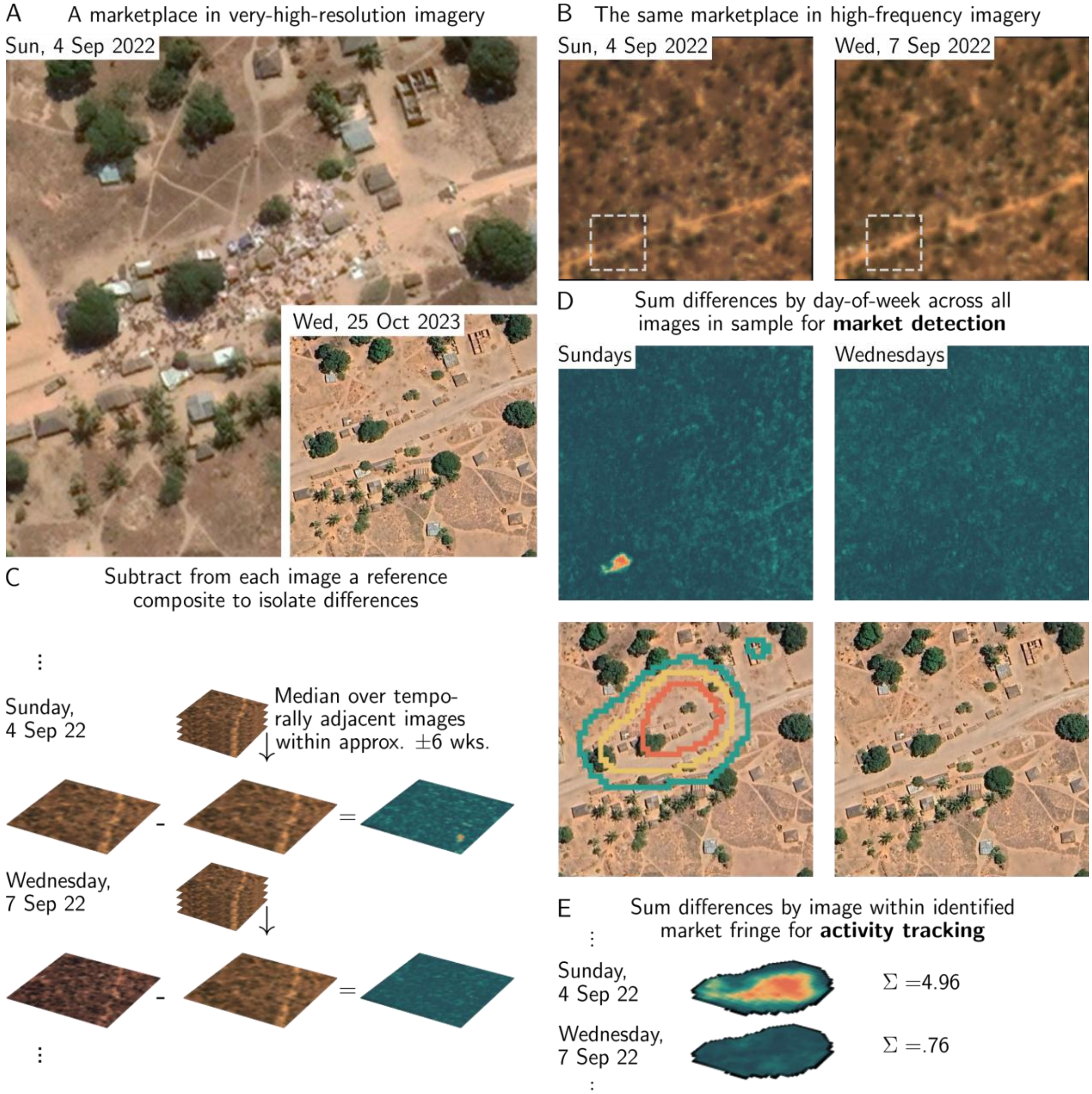

***Fig. 1: Method intuition & sequence.*** *(**A**) A marketplace in Nangata, Mozambique (14.21°S, 40.7°E) as seen in the Google Earth Archive on a local non-market day (Wednesday) and a market day (Sunday). Market activity is discernible as crowds, stalls, and vehicles in the Sunday image. (**B**) The same location on a Sunday and a Wednesday in PlanetScope imagery, with the grey box corresponding to the area shown in (A). (**C-E**) Illustration of workflow for market detection and activity tracking. Individual PlanetScope images between June 2016 and September 2024 are differenced with median composites from temporally adjacent images – approximating the*

*appearance on a non-market day – to isolate appearance differences due to market activity (C). Aggregate differences per day-of-week identify market shapes at various threshold levels (D). The high-resolution images show the area in which market shapes are detected at various threshold levels. Market activity readings within these shapes are derived by summing differences across pixels within each shape and image (E). We then select as the market area the largest shape in which the median reading across market days exceeds the 75$^{th}$ percentile across non-market days.*

Across marketplaces, market-day activity varies in its appearance compared to non-market days. Sometimes it is brighter than the otherwise bare ground, sometimes darker; vendors often use temporary structures of distinctive but locally idiosyncratic colours. We address this heterogeneity by aggregating the information in each pixel of each day-of-week difference image into a summary measure of brightness and colour differences. Specifically, we multiply the maximum absolute difference of values across the red, green, and blue bands – tracking brightness variation – with the maximum absolute difference of angles in a polar representation of the image[19] – tracking colour variation.

We identify contiguous areas where the signal from combined brightness and colour differences exceeds predefined thresholds (lower row of Fig. 1D). We show using ground-truth data that the stronger the signal is, the more likely it is that the areas we detect correspond to a periodic marketplace (see following section). While this procedure could potentially identify other periodic events like religious gatherings, we confirm below that we indeed predominantly detect marketplaces.

Given the available imagery and our choices in designing the algorithm, our approach will identify markets that meet the following criteria: i) occurring periodically (e.g., one, two or three market days per week); ii) at least partly taking place in the open with a detectable footprint of at least 50m$^2$; iii) being observable sufficiently frequently between June 2017 – when imagery becomes consistently available – and September 2024 – our cutoff date – to distinguish activity from idiosyncratic variation, including cloud-related noise; (iv) and operating around 10:30 AM local time when most imagery is captured. We discuss the implications of these criteria for our method's coverage and generalizability when we introduce the novel Ethiopian market map.

We now turn to measuring activity over time. Marketplaces typically feature a core with relatively stable activity and a fringe with seasonal or other fluctuations, but their relative positions vary across markets. We first identify each marketplace's fringe by creating a series of concentric rings originating from the part of the market with the strongest periodic signal, based on the same pre-defined thresholds that we used to delineate market outlines. Within each ring and for each image, we compute the mean of brightness and colour differences as above. We then define the fringe of each marketplace as the largest ring where the median measure across market days exceeds the $75^{th}$ percentile of non-market days. Intuitively, this should identify areas where activity does fluctuate – as opposed to more constant activity in the market's core – while at the same time limiting the influence of noise from adjacent areas that markets rarely or never expand into. For each available image, our preferred measure of activity is then the mean of the brightness and colour differences across pixels within the area constituting the outer boundary of the fringe (Fig. 1E). This measure reflects the density of structures, vehicles and crowds on market days within the typical market area, compared to non-market days. The resolution of the underlying imagery does not allow us to discern the components separately.

The derived metrics of appearance differences originate from heterogenous marketplace layouts, and their absolute values are therefore not directly comparable across marketplaces. To create an index that can be compared across marketplaces, we normalize these metrics by market and day-of-operation with their average during a specific reference year. We adjust for the varying relative frequency of images throughout the year (see Fig. 3B) so that the index corresponds to an annual average of market activity.

Detecting and tracking marketplaces requires substantial amounts of imagery as well as processing capacity. To use both imagery and processing resources efficiently, we focus our search in the exercises that follow on specific areas ('candidate locations'). In the first exercise, where we evaluate how successfully we detect marketplaces that are recorded in validation datasets, we apply our market detection approach to polygons drawn around the settlements that contain each marketplace. When we map markets in Ethiopia, we use secondary data and freely available lower-resolution imagery to progressively narrow our search to areas with a relatively high *ex-ante* likelihood of containing marketplaces, such as villages or roadsides, as opposed to unpopulated areas.

**Limited validation data suggests successful detection of markets**

We now assess the accuracy of our detection method. Using the limited available sources of validation data, we confirm that we successfully detect known marketplaces and that detected areas overwhelmingly correspond to trading locations rather than other periodic events.

To validate our detection procedure, we use secondary datasets from Kenya[20], Malawi[21], and Mozambique[22]. These datasets are uniquely useful in that they originate from exercises specifically targeting periodic marketplaces, listing the locations and days of operation for 60, 31, and 48 marketplaces, respectively, with 1.2 active market days per market on average (Fig. S1). For each listed marketplace, we outline a corresponding candidate location by manually encircling the settlement containing the marketplace. Note that these areas of interest are much larger than the markets themselves, with an average size of 181ha compared to the average detected area of the marketplaces (0.97ha).

The validation datasets let us directly test our ability to detect periodic markets where they are known to exist. However, we lack maps of otherwise comparable locations without marketplaces. To examine how frequently we erroneously detect markets, we use imagery from designated non-market days to simulate pseudo-instances of such locations. Specifically, we remove all images of designated market days for a given location from its imagery sample, replacing them with randomly sampled images from designated non-market days. The resulting set of imagery should have the same appearance on average, and importantly the same non-periodic variation in appearance, as an otherwise comparable location without a periodic market. We then calculate the share of these pseudo-locations that are erroneously identified to have a market on any day (‘false positive rate’ (FPR)). Within the original set of locations, we calculate the share of location-day tuples that are correctly detected (‘recall’) and the share of detected tuples that are designated market days for a given location (‘precision’).

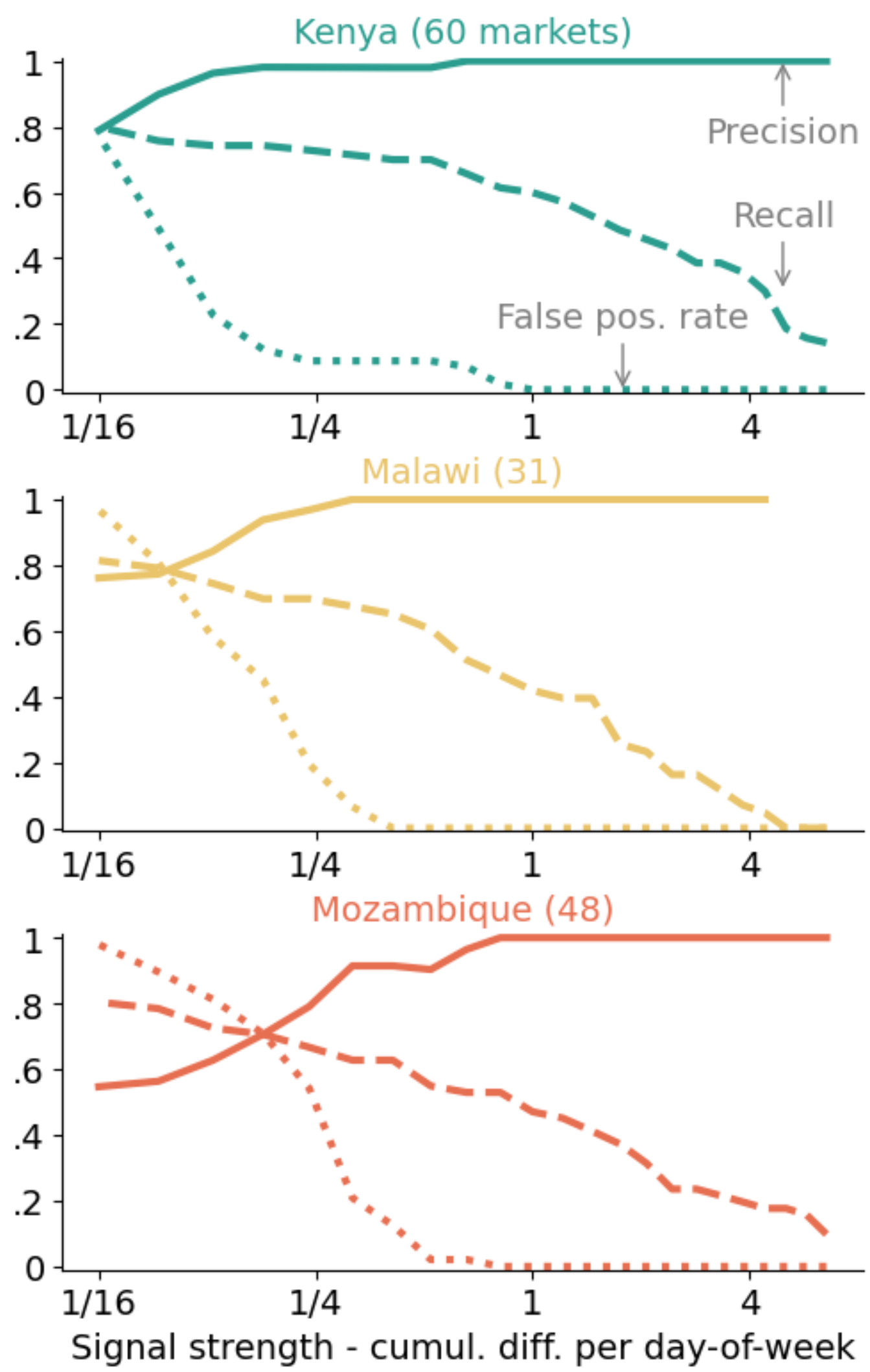

***Fig. 2: Validation of market detection against ground-based marketplace maps.*** *Precision, recall and false positive rate (the share of pseudo non-market locations for which the algorithm erroneously detects any marketplace) across market datasets from three countries. The*

*horizontal axis indicates a range of values for signal strength in each of the candidate locations that we screen.*

Despite varied economic and agro-ecological settings, we achieve precision above 90% in the sample of locations with marketplaces, and an FPR below 10% in the sample of pseudo-locations without marketplaces, across a broad range of signal strength (Fig. 2). The detected shapes are small in relation to the initial candidate locations (mean market area share of candidate location area 3.5%; median: 1.6%). This suggests that even when screening relatively large areas, appearance differences on market days will, if present, stand out clearly against other potentially present appearance changes, both periodic and non-periodic.

Recall rates decrease with signal strength and never reach 100%, driven by locations where we do not detect any activity, even at low signal strength, on designated market days. Several factors may contribute: first, although we select datasets that list marketplaces with distinct designated market days, some marketplaces may also have substantial levels of activity on other days of the week, complicating their detection. Second, some market locations, particularly in Mozambique, refer to large trees, which often serve as local gathering places. Periodic activity may indeed happen in their shade, but this would not be detectable in remotely sensed imagery. However, in these cases, any activity is likely on a substantially smaller scale than the markets we successfully detect. Finally, the datasets were collected at different times. Only markets that were active for a sufficiently large share of our imagery period will be detectable. These factors do not reflect shortcomings of our specific validation data but intrinsic challenges when studying such a dynamic and diverse phenomenon. No current approach can detect and monitor economic hubs of *all* forms and sizes. Our approach allows detection and monitoring of a consistently defined group.

Regarding potentially misidentified marketplaces, our method demonstrates high precision and a low false positive rate at high signal strength. We can investigate the types of phenomena that induce false positives by overlaying detected shapes onto VHR imagery. At a signal strength of 1 in Fig. 2, we identify two cases in Kenya where livestock pens adjacent to marketplaces are detected on days immediately before or after designated market days – likely livestock brought to the market in advance or awaiting transport after sale – as well as one possible instance of a repeated gathering in front of a public building.

**A map of periodic markets in Ethiopia**

We focus the remainder of the paper on a novel panel of marketplace activity in Ethiopia. This dataset allows us to characterise where market activity may be informative about economic conditions and to examine the sensitivity of measured activity, as well as to illustrate an application of our approach to tracking disruptions caused by recent conflict.

To search efficiently for markets, we initially manually drew 450 polygons around settlements and road intersections, using VHR imagery, to delineate areas where marketplaces were likely to exist. We screened those candidate locations using PlanetScope imagery following the approach shown in Fig. 1B-D. Building on this pilot, we then developed a more scalable approach to identifying candidate locations: first, we focus our attention to where markets can reasonably be expected by identifying areas within 300m of roads recorded in OpenStreetMap or where the countrywide Global Human Settlement Layer[23] suggests built-up area (Fig. S2A-C). This excludes vast areas that are uninhabited or solely used for agriculture. We also exclude Addis Ababa: markets in very large cities are less likely to be periodic as higher aggregate economic activity can sustain permanent, more formal trading systems[1], and there is higher noise and more confounders for our approach, such as periodically-filling parking lots. Limiting to settlements and areas near roads and excluding Addis Ababa narrows our search to 4.5% of Ethiopia's landmass. Second, we replicate the approach from Fig. 1B-D with Sentinel-2 imagery, which comes at a lower resolution and frequency, but allows us to screen the entire remaining area for markets' characteristic periodic appearance changes at very low computational cost. This Sentinel-2-based screening procedure returns pixel-wise measures of periodic appearance changes ('S2 screening signal'). We identify clusters of elevated S2 screening signal with a radius of 250 meters (mean size: 24ha), and assign to each cluster its maximum S2 screening signal (Fig. S2D). Ranking the resulting candidate locations by that signal (Fig. S3B,C), we then screen locations in decreasing order using the higher-resolution and higher-frequency PlanetScope imagery. Locations with lower S2 screening signal are less likely to be marketplaces, so we conclude our screening when it becomes uneconomical, i.e., when the share of candidate locations returning a marketplace falls below 10% (Fig. S3D). We also identify locations labelled 'marketplace' in OpenStreetMap that we did not identify either manually or with Sentinel-2 imagery (93 locations out of 421 listed in OSM in total). In sum, we screen 0.093% of Ethiopia's landmass across 4,562 candidate locations with PlanetScope imagery. The output of this process is a set of outlines of areas with periodic appearance changes of varying intensity ('signal strength') which may include artefacts from noise, or periodic events other than market activity as well as marketplaces.

The final step makes use of the fact that marketplaces typically have intuitive layouts and locations, such as along roads or on village squares. We visually evaluate the outlines returned by the PlanetScope-based screening procedure by superimposing them onto VHR imagery. We retain locations that resemble marketplaces regardless of signal strength, and we exclude locations that may exhibit elevated appearance differences but which are apparently not marketplaces (see Figs. S3A-B, S4; the online replication package also contains examples of outlines we labelled as not representing marketplaces.) The latter category includes parking lots and storage areas around industrial operations, as well as artefacts resulting from noise. Because the number of features that exhibit brightness variations across imagery -- such as reflective roofs or water surfaces – is large, some small share of these will appear periodic by chance. These are, however, straightforward to screen out in VHR imagery. At the lowest level of signal strength returned by the procedure, no outlines appear to correspond to marketplaces (Fig. S3B), suggesting that our procedure is unlikely to wrongly exclude markets with activity that we should be able to detect with our approach.

Contrasting our manual examination of detected shapes to a simple decision rule based on an accuracy-maximizing threshold of signal strength, we find that 96.1% of detections with signal strength above the threshold are ultimately classified as marketplaces, while 20.4% of all eventually confirmed marketplaces would not have been detected at the accuracy-maximizing threshold (Fig. S3C). The manual examination therefore mostly expands the spatial coverage of the data relative to the simpler decision rule. Even without manual screening, precision is high.

Following this procedure, we detect 1,776 markets with 1.10 days of operation and covering .72 hectares on average (Fig. 3A; see Fig. S5 for example detections. The online replication package contains an atlas of all detected markets). When we detect multiple market days at the same marketplace, they are typically scheduled three days apart – as is characteristic among periodic markets[1] – and their centroids lie, at the median, only 19 metres apart.

The detected markets are spread widely across the populated parts of the country outside the major cities: we estimate that 69.7% of the country's population (73.3% of the population outside Addis Ababa) live within 10km of a marketplace (Fig. 3B). With markets distributed throughout most of the country, the market activity measures will be most useful for economic monitoring in rural areas – home to 77% of the population[24] – where alternative sources of data are scarce relative to population density. There are also large parts of the country where we do not detect many periodic markets, especially in the eastern Afar and Somali regions, despite having identified candidate locations there as well. This apparent lower prevalence of weekly markets likely reflects the regions' lower population densities, as well as their pastoralist and semi-nomadic character.

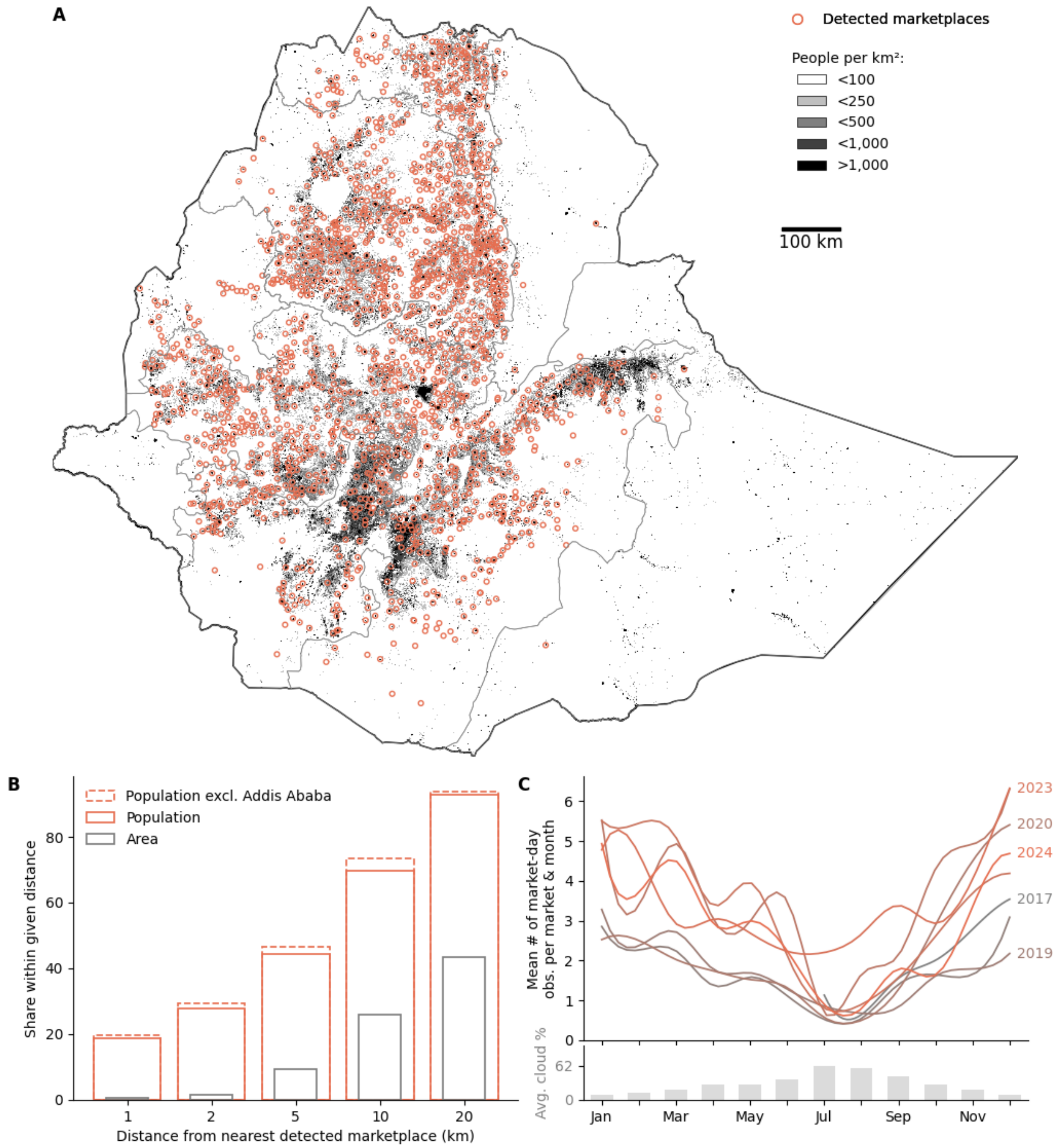

***Fig 3. Spatial and temporal coverage of a novel marketplace map of Ethiopia.*** *(A) The map shows the marketplaces detected through a country-wide screening using our method, superimposed on population density data*[3] *and regional borders. Addis Ababa was excluded from the screening. (B) Shares of population (including and excluding Addis Ababa) and country area lying within various distances from detected marketplaces. (C) Smoothed averages of the count of market-day activity measures across the Ethiopian sample for each year in our panel. Availability is driven by satellite launch dates, seasonal cloud cover, and the number of active*

*market days. The bottom panel shows the average share of pixels classified as cloudy in Sentinel-2 imagery between 2018-2024 across our detected marketplaces.*

The underlying logic of our approach defines the type of market we detect, among the universe of phenomena that might colloquially be called “markets”. To be observable, trading must occur periodically, outdoors, and around 10:45am local time, when the bulk of imagery is acquired. Fig. 3A illustrates that we detect markets meeting these criteria over a wide geographical range. There is an element of fortuitous coincidence here in that the features that make markets detectable in satellite imagery are also common for entirely independent reasons. Midday gatherings accommodate participants traveling from surrounding areas. Convening on fixed, publicly known days increases the likelihood of finding trading partners in otherwise thin markets, allows time between market days for goods production, and accommodates traders who circulate between markets[1]. Periodic markets also, by nature, operate in the open air, with temporary stalls that can be quickly assembled and disassembled. This pattern holds even as markets begin to formalize with more permanent structures: some detected areas of periodic activity encircle fixed structures, illustrating that even when markets are partly covered, activity spills into surrounding areas where it is detectable (Fig. S5 and atlas in replication package).

A possible confounder we did not yet discuss could be religious gatherings that also follow a weekly cycle. However, only one of the 1,776 detected locations lies within 50 metres of a religious building registered in OpenStreetMap and has a signal we detect on days of worship (Tab. S1).

Comparing our satellite-derived map to existing sources, we find that although our approach detects many more marketplaces than currently registered in OpenStreetMap (N = 421, compared to N = 1776 using our approach), 71% of OSM-listed marketplaces outside Addis Ababa lie within 50 meters of a marketplace we detect (Fig. S6C). The remainder are frequently located in larger towns and are likely to be daily markets, which our approach would not detect. We detect somewhat fewer marketplaces than recorded in the last available census of marketplaces from 2007[25] (n=2,627). The inclusion criteria for the census are unknown, and many locations appear to be registered wrongly, with centroids that lie in dense vegetation or distant from population centres. In fact, a 3.5 times larger share of the population lives within 1km of our detected marketplaces than within the same distance of the census marketplaces, despite the smaller number of detected marketplaces compared to the census (Fig. S6A). Furthermore, 34% of census marketplaces are in very sparsely-populated areas, with fewer than 100 people estimated to live within 200m, while only 3.2% of detected marketplaces are in similarly sparsely-populated locations (Fig. S6B). While our screening process means that we would not detect markets in very sparsely populated locations away from roads, these statistics suggest that unless the census is recording a completely different phenomenon, the locations reported in the census may be inaccurate.

We also compare our map with data from the Living Standards Measurement Surveys in 2015, 2018 and 2021, which asked informants whether there existed a “large weekly market in their community”. Geocoordinates for LSMS enumeration areas are randomly offset to anonymize respondents, precluding an exact spatial match between these data and the detected markets. Still, respondents that report having a large weekly market within 5km are systematically closer to our detected markets than locations that report not having a market within 5km ($p<0.001$, from a Kolmogorov-Smirnov test; Fig. S7)

Finally, we compare our detected markets against raster data of light emissions at night, the most commonly-used remotely-sensed proxy for economic activity. 63% of our markets are found in locations without any measurable light emissions between 2022 and 2024 (Fig. S8). Consistent with other analyses that show that nightlights are not informative about levels or changes in economic conditions in poor rural areas[13,15], this pattern of results confirms that nightlights would not be able to track changes in economic conditions at these locations.

Equipped with this map of marketplaces, we can then derive the activity measure underlying the exercises in the remainder of the paper. The measure’s availability over time primarily depends on the number of satellites in orbit, which increased after 2020, as well as seasonal cloud cover (Fig 3C).

**Measured market activity varies with agricultural calendar and weather**

Fig. 4A shows data for one marketplace over a three-week period. For this and the following exercises, we normalize the activity measure for each market by first subtracting the non-market day average in 2018 and then dividing by the market-day average from the same year and multiplying with 100. This procedure ensures that 100 corresponds to the average measured activity on market days during the reference year of 2018, and 0 corresponds to the average measured activity on non-market days. Importantly, note that both market-day and non-market day activity measures are scaled identically. Despite this identical rescaling, and as also shown in the illustration of the approach in Fig. 1E, activity readings have markedly different distributions on designated market days and non-market days. This level difference, as well as the clear separation of their ranges, would not appear if the raw activity readings did not clearly distinguish the extent of attendance on market days from their appearance on non-market days. Fig. S9 confirms that measures differ significantly between market days and non-market days across most detected marketplaces.

Beyond registering market attendance, changes in market activity are potentially useful as a proxy for changes in local economic activity. When incomes are higher, more people gather in the places where goods and services are available for purchase. Similarly, when more goods are being produced locally, more producers will also be drawn to those same locations, where buyers are to be found. Since we lack market-level measures of attendance at sufficient temporal and spatial density for direct validation, we instead evaluate whether our activity measure responds to well-established locally relevant drivers of supply and demand.

Given rural areas' reliance on rain-fed agriculture[24], we expect seasonal patterns of market activity to reflect seasonal patterns of rainfall[8], peaking around and after harvest, when crops are sold.

Fig. 4B illustrates this pattern for the 50 markets detected in the Central zone of the Tigray region. Activity is lowest early in the growing season – a period often referred to as the lean season[26] – and rises sharply towards the harvest season. The sharp and unseasonal decrease in market activity in April 2020 coincides with the imposition of restrictions in response to the Covid-19 pandemic[27].

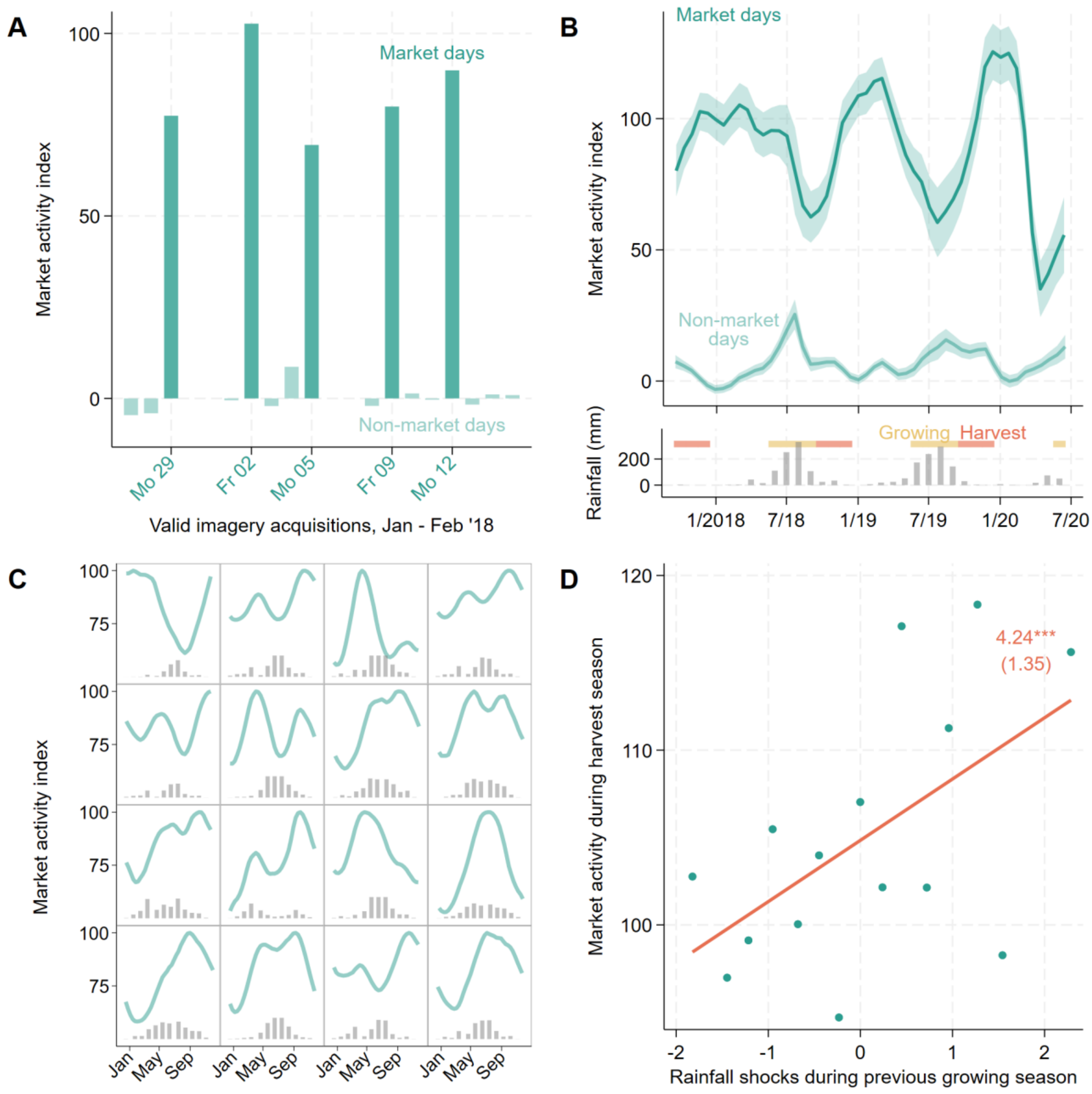

***Fig. 4: Validation of activity tracking with Ethiopian sample markets.*** *(**A**) Readings for market at 9.29°N, 38.56°E, where 100 (0) represents the average market activity in 2018 on market days*

*(non-market days) from an interpolation of actual readings across the whole year to account for seasonal variation in image availability. (**B**) Smoothed averages and 95%-confidence intervals of activity readings for 50 markets in Tigray's central zone, 7/2017-6/2020. The bottom panel shows average monthly rainfall across the marketplaces in the sample as measured in* [29]*. Growing and harvest season taken from* [28]. *(**C)** Smoothed 2018 averages for the 16 zones with the highest number of marketplaces, with grey bars indicating monthly rainfall. The top left panel corresponds to the zone shown in (B). (**D**) Binned scatter plot across zones, excluding Afar and Somali regions as pastoral areas with different production systems. Horizontal axis shows standardised rainfall shocks during June-August relative to 1998-2024 average; vertical axis shows average activity across markets (100 = 2018 average) for October-December in 2017-2024. Orange line represents linear fit from a regression with fixed effects for each zone. Standard errors clustered at the zone level.*

One might be concerned that these seasonal patterns simply reflect environmental changes, such as varying vegetation or cloud cover. We verify that this is not the case by comparing variation in market day activity to measured variation in non-market day activity. The latter also varies over time, but with markedly different seasonal patterns, likely reflecting residual weather-related noise (e.g., clouds and their shadows appearing as brightness and colour differences). Importantly, non-market day seasonal fluctuations are substantially smaller in magnitude, suggesting that the observed seasonality in market-day measures reflects genuine changes in market activity (Fig. 4B). Indeed, the apparent increase in non-market day activity during the rain season most likely reflects the effects of noise from cloud cover, in turn suggesting that the observed rain season decline in market activity understates the true decline.

The zone shown in Fig. 4B lies in a region with a strong unimodal rainfall cycle. Fig. 4C shows how the seasonal pattern of market activity differs across zones with different rainfall patterns, which in Ethiopia vary between unimodal and bimodal[28].

The seasonality in market activity could, in principle, reflect people's taste for visiting the market at specific times, e.g., when it is not raining, rather than indicating changes in economic activity. Beyond the intra-annual variation in Figs. 4B&C, however, we also see differences in activity levels across years. One potential driver of inter-annual variation could be growing conditions for agricultural crops, which determine the abundance of harvests and, subsequently, the amount of crops households sell as well as their demand for consumer products sold at marketplaces. Harvest-season market activity should therefore increase in response to more abundant rainfall. Fig. 4D confirms this expectation for the 61 administrative zones during 2017-2024. The positive association we observe – market activity increasing by 4.24 index points for each one SD increase in rainfall – is likely an underestimate of the true relationship: we focus on the main growing and harvest seasons across Ethiopia, but their precise onsets vary across space and time. Also, normal patterns of seasonality were disrupted in many zones during the study period due to conflict or the Covid-19 pandemic.

To provide a quantitative sense of this effect, we can compare its size to the typical reductions in market activity observed during the lean season (Tab. S2). Comparing the least busy quartile of month per administrative zone with the middle two quartiles, we find that activity on average decreases by 17.7 index points.

Taken together, these exercises highlight the ability of our remotely-sensed measure to detect meaningful changes in a potential proxy for economic conditions, including over short time periods such as during the 2020 lockdowns or when activity picks up around local harvests. As with most other metrics, whether we detect a signal depends on its strength relative to background noise. In our data, noise arises from other sources of variation in images, such as haze, changing acquisition times, or lateral distortions between images[30]. Using the standard deviation of month-to-month differences across various sample definitions, we can estimate the sample size needed to statistically detect differences of a given magnitude (Fig. S10). Statistical power is higher with larger samples of markets or over longer time periods. For example, a year-on-year difference of 4 index points can be detected with 80% power in a sample of just six marketplaces, while detecting a month-on-month difference of the same magnitude requires a sample of around 60 marketplaces.

**Market monitoring during a humanitarian crisis**

Poverty is increasingly concentrated in fragile regions[31]. Fragility not only affects economic outcomes, but also our ability to monitor them, making it costly, or impossible, to collect field data or maintain consistent administrative records[9]. Our method has particular potential to remotely detect, quantify and anticipate disruptions in conflict-affected regions, when other sources of information are sparse.

We use our data to track the consequences of the recent episodes of conflict in Ethiopia during which period widespread instability hindered access to reliable information on economic conditions[32]. Starting in November 2020, civil war broke out in the Tigray region and the following years saw fighting and instability across large parts of Ethiopia. We combine detailed geo-referenced counts of conflict events[33] with our panel of market activity. Across the three regions with most recorded events, periods with elevated conflict generally coincide with lower market activity (Fig. 5A). We also observe sharp decreases in market activity around the imposition of Covid-19 restrictions in the second quarter of 2020, with decreases being most pronounced in the Tigray region.

It is important to note that conflict could have resulted in the reallocation of trading activity away from periodic markets to smaller, even less formal modes, which would be unobservable in our data. In that case, the observed decreases in market activity would represent an upper bound on decreases in activity more broadly. It is similarly also possible that other types of economic activity are disrupted to an even greater extent, if markets remain an essential source of subsistence goods, in which case the observed decreases would represent a lower bound on decreases of activity. In either case, the sign and the timing of the effects we observe remains informative about the spatial extent and duration of conflict-related disruptions.

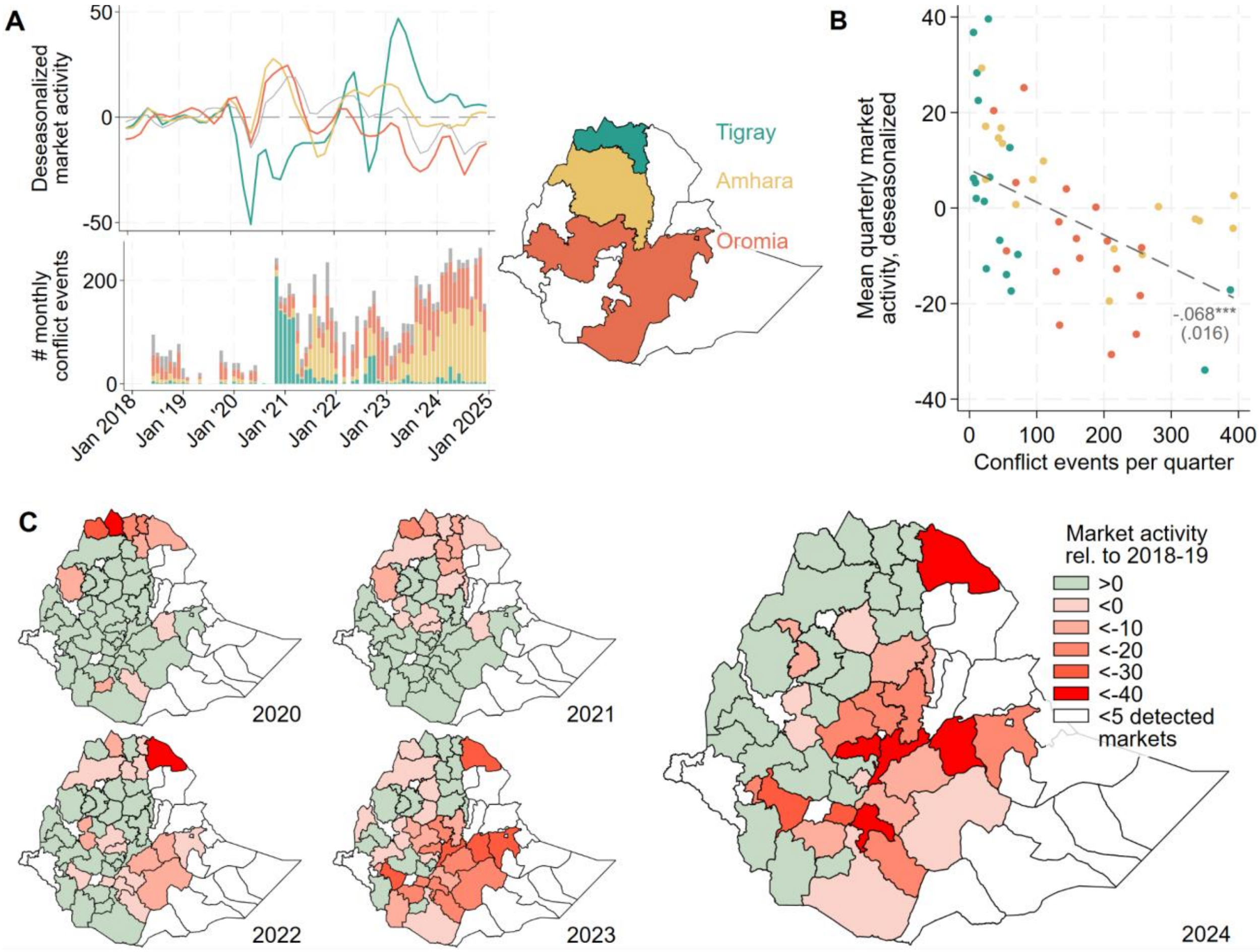

***Fig. 5: Application of approach to country-wide monitoring in Ethiopia, 2018-24.*** *(A) Activity readings on market days across markets in three Ethiopian regions, deseasonalised according to their 2018-19 month-of-year averages, along with counts of conflict events by region[33]. Grey lines and bars correspond to areas outside the three regions highlighted in the map. (B) Counts of conflict events and mean measures from (A) by region, quarter and year, starting from Q4 2020. The dashed line represents a linear fit, with the associated coefficient estimate from a regression with heteroskedasticity-robust standard errors. (C) Administrative zones shaded by average market activity measures from (A) per year. Regions with less than 5 detected markets are left empty. Administrative borders are taken from the FAO Global Administrative Unit Layers dataset, with some smaller neighbouring zones merged within the same region for easier interpretability.*

We confirm the visual evidence of a negative association between conflict and market activity in a regression (Figs. 5B, S11). The panel also highlights substantial variation in market activity for a given level of conflict. Although some of this variation may reflect noise, it also suggests that remotely-sensed market activity measures could contain information on local impacts beyond the more readily available conflict records.

Beyond regional correlations, our data also allows for finer – including arbitrary – spatial disaggregation: such an exercise reveals substantial heterogeneity in market activity changes across administrative zones (Fig. 5C). As with the regional level in Fig. 5A, the magnitudes here are substantial – a 30 index-point decrease as observed in many zones exceeds the seasonal swings estimated in Tab. S2. This indicates that in some years and regions, market activity on average did not exceed levels typically observed during the lean season.

**Discussion**

Anecdotally, periodic markets appear to be widespread beyond Ethiopia and the three countries in our validation data. However, no consistent map of periodic marketplaces across countries currently exists. Detecting and monitoring rural marketplaces with satellite imagery provides an unprecedented opportunity to study this fundamental economic institution more broadly. The ability to track rural market activity also makes it feasible to monitor a potential proxy for economic conditions in remote or conflict-affected regions in near-real time and at high frequency.

Our approach has potentially broad applications, especially in settings where other measures of economic conditions are sparse. A future application of our approach could be to complement existing early-warning systems that aim to detect humanitarian crises in time to intervene or prepare a response. By nature, these are information-sparse environments, in which policy-makers have to allocate resources with limited information. Measures of weather conditions, such as drought, can provide information about exposure to natural hazards, but do not provide information about the consequences for livelihoods, which may vary across contexts depending on the resilience of support systems. Other remotely-sensed metrics, such as nightlights, respond only sluggishly to changes in economic conditions. Phone surveys can gather information quickly, but mobile phone networks or power networks may fail during severe crises. In these information-sparse environments, remotely-sensed market activity can potentially expand the set of tools available to policy-makers to detect emerging crises by providing timely, location-specific information about economic activity, independent of ground conditions. Researchers may also be able to utilize these data to better understand the economic impacts of extreme events or policy changes in remote rural areas.

There are several limitations to our approach. First, limited ground-based panels of economic activity in general–and market attendance specifically–inhibit our ability to validate the activity measure against fully comparable ground truth data. This prevents us from, for example, examining whether the elasticity between market activity and local economic conditions may be non-linear. Here, correlating our activity measure with other high-frequency measures such as mobile phone data appears promising. This elasticity might furthermore change over time as economies develop and formalize. While the activity measures can be constructed from 2017, further analyses could substitute space for time and carefully compare across places at different stages of economic development.

Second, elasticities could differ across marketplaces if, for example, our measure is more sensitive for markets that are entirely open-air, compared to partially-covered ones, or in places where alternatives to trade at marketplaces are more readily available, for example, depending on the local viability of reverting to subsistence agriculture. Such heterogeneity would complicate comparisons across markets, or groups thereof, as a unit change in measured activity would correspond to differently sized changes on the ground for different groups. There are three potential strategies that may help researchers avoid drawing erroneous conclusions. First, making comparisons within units – such as individual marketplaces or groups thereof – such as in Fig. 4B will prevent bias from cross-market heterogeneity. Second, the signs of changes – e.g., the share of markets in an area where activity is lower than during some reference period – may be more reliable than the magnitudes of changes. Finally, comparisons across marketplaces are likely to be more informative when the marketplaces are similar. Researchers may be able to use measurable features of the markets – such as their size and shape, their position relative to infrastructure, and their seasonality – to shed light on the degree of comparability of markets in different comparison groups.

Another limitation to the accuracy of measured market activity originates from the quality of the underlying imagery. While PlanetScope's resolution, frequency, and global coverage uniquely enable our approach, challenges remain: spatial misalignment between images, different spectral response functions between sensors, and imperfect cloud masking introduce idiosyncratic noise. In particular, the presence of cloud cover substantially affects market activity signals, implying that we may not detect markets that take place in areas with high and persistent cloud cover, and signals from these areas will be noisier. For now, this may be an important constraint on future applications to such places, limiting the geographies in which market activity may be useful as a proxy for economic activity. Future improvements in sensor design may, however, alleviate some of these constraints.

Finally, we designed our method to balance potential transferability across heterogeneous climatic and geographical contexts with the need to maintain reasonable computational loads. This may lead us to miss important local heterogeneities. For example, cattle markets appear characteristically dark in the satellite imagery, while most other markets appear relatively colourful, a difference we ignore by relying on absolute differences. Our approach could, however, be locally tailored to distinguish such kinds of markets – or areas within marketplaces – especially when only a small region or a specific period is of interest. Also, to reduce computational load and the costs of imagery acquisition, we focus our search on areas with either permanent settlements or transport infrastructure networks. If markets gather away from such areas, our current approach would not detect them. Extending screening to these areas would be straightforward but expensive. Furthermore, while we have focused on markets occurring at weekly frequencies, our methodology can straightforwardly be adapted to settings where markets occur at other intervals, as long as they occur regularly, in the open air, and not on the majority of days. The latter is necessary so that seasonal changes in the appearance of market areas that occur independent of market activity can be differenced out with a reference composite resembling a non-market day.

Generally, however, our method, in combination with the global availability of the required imagery, opens the door to a deeper understanding of a universal institution that is closely linked to people’s livelihoods in remote rural regions.

**Funding:** We gratefully acknowledge support from Riksbankens Jubileumsfond Infrastructure for Research Grant IN22-0041, STEG Small Research Grant 64, and the IGC project grant MOZ-22258. This work was also supported by an FSE postdoctoral fellowship to TC.

**Supplementary Materials**

**Materials and Methods**

**Market detection**

The following paragraphs provide more detail on how we screen the country for candidate locations, and how we get from candidate locations to, if a market is confirmed, a shape outlining market areas and a time series of activity measures. Our design choices reflect the goal of transferability across geographical contexts, the limitations of sparse ground truth data, and the need to maintain reasonable computational loads when applying the approach at scale.

**Identifying candidate locations.** While the PlanetScope imagery would in principle allow for a true global screening for periodic appearance differences, such an approach would be prohibitively costly in terms of the necessary imagery and computation. We therefore began by identifying likely market locations from visual interpretation of very high-resolution imagery. Learning from these locations then allowed us to configure a country-wide screening mechanism building on mid-resolution Sentinel-2 imagery as well as roads recorded in OpenStreetMap and gridded data on built-up area[23] to exclude unpopulated areas. Specifically, we identify all areas that lie within 300 meters of a primary, secondary or tertiary road, as well as those where the gridded data indicates sufficiently densely clustered built-up pixels (road surfaces or buildings) according to an aggregator function. We choose the function empirically based on visual assessments of the data superimposed on VHR imagery (Fig. S2A-C).

For each of the manually identified locations, we drew polygons around areas that could conceivably host a market – including built-up and adjacent areas but excluding waterbodies, forests and agricultural land. For the Sentinel-2 derived candidate locations, we buffered locations of elevated periodic appearance differences by 250 meters (Fig. S2D&F). For the OSM markets, we similarly buffered their centroids by 250 meters.

**Selecting imagery.** Given a candidate location, we identify all PlanetScope Scenes that i) intersect the area of interest and ii) have less than 50 percent cloud cover, according to the provider, across the whole image tile (287 to 637km$^2$, depending on the imagery generation). We restricted our search to RGB imagery acquired after June 1, 2016 and before September 30, 2024 and downloaded orthorectified surface reflectance analytical data (Level 3B products)[34]. For a typical location this returns an initial sample of about 2,200 images. During the downloading stage, we make use of the provider's pipeline[34] to spatially align images using a post-2020 image with as little cloud as possible as an anchor. This reduces distortions of the imagery created by varying sensor angles during acquisition.

We then process the imagery in Google Earth Engine. We first aim to identify whether a given candidate location indeed has areas of periodic appearance changes. To this end, we exclude outliers more aggressively than in the ensuing activity measurement, as in the latter, even noisier measurements may be valuable. We begin by masking each downloaded image with the provider's usable data mask ('UDM2') and filter the image collections to only include images covering at least 20 percent of the candidate location and being at least 80 percent shadow-, haze- and cloud-free across the whole image tile according to the imagery metadata (discarding about 10 percent of the initial sample). We furthermore exclude images that are acquired more than 30 minutes before or after the median acquisition time for each location to reduce variance in lighting conditions (a further 20 percent of the initial sample). We also filter out images with faulty colour representations, defined as images where the mean across pixel values within any of the red, green and blue bands is more than two standard deviations away from the mean of that statistic across all images (a further 5 percent of the initial sample).

**Imagery harmonization.** A complication stems from the two generations of satellites deployed by the provider during our sample period. Each generation operates with its own spectral response function, leading to the same colour on the ground being represented as different RGB values in images from either generation. If not properly addressed, these colour differences could be confused with changes in market activity. During the downloading stage, we follow the provider's recommendation[30] and spectrally align each image with Sentinel-2 imagery, harmonising colour representations in our imagery both across the two generations, and within each generation across individual satellites. Even after this adjustment, however, we observe that ranges of the RGB band values differ between the two imagery generations for the same location. For each location, we therefore create calendar-monthly target composites from images from the newer generation, to which we spectrally match images from both the older and newer generation. Specifically, we calculate 256 quantiles for each band in both individual old-generation images and the new-generation target composites and subsequently align distributions by scaling each quantile in the former to the value of the same quantile in the latter.

**Reference composites.** We then proceed to construct reference composites to serve as a comparison for each individual image. Our goal is a reference composite that for any given location and date resembles the appearance of that location when there is no market taking place. We therefore sample images that are temporally close – within $\pm 6$ weeks of each image's acquisition date – and maximise contrast with potential market days by excluding images taken on the same day-of-week. From this initial sample, we select three copies of each of the two closest instances of every different day-of-week, two copies each of the next two closest instances and one copy each of the next two closest, giving us a maximum sample of

$$6\ days\ of\ week \times\ 2\ \times\ (3+2+1)\ \frac{images}{days\ of\ week} = 72\ images$$

for each image's reference composite. Parts of each image in this series may be masked out because of cloud cover: therefore, the effective number of images available for aggregation in each pixel will be lower than 72. We then construct an 'interval mean'-composite from all pixels falling between the $40^{th}$ and $60^{th}$ percentile per band in the sample. In contrast to a canonical median composite, we found from inspection of the resulting difference images that this enlarged window helps address low-level noise coming from residual spatial misalignment between images.

This way of selecting imagery for the reference composites performs satisfactorily in areas where clouds and their shadows are seasonal or generally cover large areas. Visual inspection, however, revealed that despite the multiple pre-processing steps, composites were often still noisy in areas where clouds typically occur in small patches of cumulus clouds, such as along tropical coasts. Here, a larger share of pixels in the series is masked, and unmasked pixels are more likely to be affected by noise from shadows or remaining cloud fragments. To address this, we identify in each reference composite from the first iteration those pixels where the reference composite includes fewer than half the target number of images to aggregate over. For these, we expand the sample by the next six temporally closest images per day-of-week. While this expansion reduces temporal proximity, climates with such cloud patterns tend to see less seasonality. When seasonality is low, composites from more temporally distant images can still provide a good approximation of a non-market day's appearance.

**Defining appearance changes.** By inspecting images across geographical contexts, we furthermore find that brightness variation alone – as represented by differences within each of the red, green or blue colour bands – does not capture the colour variation associated with market activity relative to varying ground colours. We therefore calculate for each pixel a polar representation of its colour according to the following formulas:

$$\theta_1 = arctan(\frac{\sqrt{red^2 + green^2}}{blue}); \quad \theta_2 = arctan(\frac{green}{blue})$$

$\theta_1$ and $\theta_2$ can be understood as angles between the individual bands identifying colours. We then match each individual image to its reference composite and calculate the absolute difference within each of the three colour bands as well as their angular descriptors $\theta_1$ and $\theta_2$. As our goal is to identify areas with periodic changes, we then calculate another interval-mean composite between the 40$^{th}$ and 60$^{th}$ percentile of these difference images sorted by day-of-week and converted to absolute values. In this representation, a pixel with a large value is one whose brightness and color (red, green, blue, $\theta_1$, $\theta_2$) are frequently different on a specific day-of-week compared to other days.

We want to allow for both brightness and colour variation to indicate busier markets, and therefore multiply the maximum absolute difference across the RGB bands with the maximum absolute difference across the $\theta_1$, $\theta_2$ bands. In this representation, a high value indicates a pixel where brightness and colour frequently differ from the reference composite on the specific day.

**Smoothing.** Equipped with this representation of periodic differences across the extent of the candidate location – one difference image for each day of the week –, we then aim to extract areas with relatively high differences on any given day. We found that brightly reflecting surfaces, especially rooftops, regularly create artefacts in the difference images as their colour can change drastically when illuminated from different angles. These artefacts are sufficiently large and frequent to not be entirely addressed by our other averaging steps. To reduce their influence, we perform two additional cleaning steps: first, we smooth out local outliers within the difference image using a 5-by-5 pixel median kernel. Second, since noise from bright rooftops is presumably uniformly distributed across days of the week – while the market signal is not – we subtract from each individual difference image an average of all other days of the week.

**Extracting contiguous areas of appearance changes**. We then turn the cleaned pixel-level representations of periodic differences into contiguous areas. Noting that they lack a natural unit, we observe typical ranges of values across the derived difference representations and define, based on inspection, a range of geometrically increasing threshold values as

$$\left\{\left(\frac{t}{20}\right)^4 | t = 9,10,11,\ldots,34\right\}.$$

We then extract all contiguous areas above each threshold, discarding those smaller than 50m$^2$ as these typically represent noise. Furthermore, we aim to eliminate spatial outliers – shapes that are identified far away from the main detected area in each location. For this, we identify for each candidate location the day-of-week and shape that is associated with the highest threshold value (a 'peak' in the difference image) and select for the same day the largest shape encircling this 'peak'. We discard all shapes that do not intersect with this encompassing shape. This filtering step assumes that the area with the strongest signal represents the core area of a potential market and essentially restricts other possible market areas to be in the vicinity of the core.

**Manual assessment.** The procedure described above returns shapes that may or may not represent marketplaces. In order to have our market map be as accurate as possible, we therefore manually examine the detected shapes superimposed on VHR imagery. We here use ESRI's satellite basemap which features images typically taken during the last three years. For some candidate locations, the detected shapes lie in areas where the VHR imagery shows a market day. Such instances allow us to learn about the typical appearance of marketplaces, even if the VHR imagery is not taken on a market day. Visual clues include village squares surrounded by buildings or wooden structures that are covered with tarps on market days.

This procedure allows us to screen out detections that are not actually representing marketplaces, but other phenomena such as periodically-filling parking lots, or noise. We perform this examination on all detected shapes in descending order of signal strength until for a given level we no longer detect any false negatives.

**Activity tracking**

Equipped with outlines of market extents under various threshold values, we turn to constructing measures of market activity. We here use all available images, including those on non-market days and those we discarded because of heavy cloud cover across the whole image tile earlier. The activity tracking follows much of the same logic as the market detection. We again construct temporally-relevant reference composites from relatively clean images, now also including images that cover less than 20 percent of the overall candidate location, but at least 10 percent of the detected market area. We note here that to apply the method in real-time, the reference composite for acquisitions within the preceding six weeks would need to be constructed slightly differently – i.e., using only imagery from the previous weeks – rather than the symmetric approach we use retrospectively here. This would introduce some additional noise as the images underlying the composites would then be less temporally proximate.

**Smoothing.** In the resulting difference images, we reduce noise from outlier pixels using a 3-by-3 pixel median kernel. For each image, we then derive the sum across pixels of the product of absolute brightness and colour differences within each of the previously detected shapes, as well as their intersections.

**Identifying market areas.** The intersections – typically resulting in ring-like shapes – allow us to identify areas where the market typically expands or contracts. Specifically, we rank the rings by the threshold value that defines them – sorting from rings nearest the core of the market to rings further away – and construct for each ring the median activity reading on market days, as well as the $75^{th}$ percentile reading on non-market days. We then define as the market fringe the smallest ring where median market day readings still exceed the $75^{th}$ percentile of non-market day readings. The outer boundary of this ring defines the limits of the market area for which we construct the activity measures. To avoid very thin rings, we restrict the collection to intersections where the smaller of the areas in the intersection is less than 70 percent the size of the larger area.

The above procedure returns two outputs: First, the shapes of the identified market areas for each market day, and, second, a table where rows identify summed deviations in a given image and the identified preferred market extent.

**Details of validation exercises and respective data collections**

**Figure 2.** Figure 2 uses three sets of market maps for validation, each of them varying not just by country, but also by how they were collected and their intended purpose. The Kenyan dataset represents the marketplaces sampled for a study of trading behaviour[20] and was generously shared by the authors. The authors reported having consulted with local government authorities for a market map but ultimately resorted to conducting their own, ground-based mapping.

The Malawian dataset originates from a policy report[21] and does not provide further detail about how they listed or selected marketplaces. However, we note that most locations represented in the map fall into relatively large settlements, such as district capitals.

The Mozambican dataset represents original data collected by the authors[22]. As part of a larger study, enumerators were tasked with travelling through districts in Nampula and Zambézia provinces along predefined routes. Market locations were identified from informant interviews as well as direct observation by enumerators. The sample used for the validation exercise includes only those marketplaces that enumerators found to be in operation during their surveys in 2021-23 and that had, according to local informants, a distinct market day. Enumerators were also tasked with photographing these locations, allowing us to identify false negatives.

All validation maps come in the form of point coordinates either from the marketplace itself or the surrounding village. To mimic a situation where market locations are not known ex-ante, we manually draw polygons around the entire settlement containing the marketplace, as identified from the Google Maps basemap, and subsequently apply the detection algorithm to these polygons.

Across datasets, we repeatedly detect areas at high threshold levels on days where the dataset does not indicate a market day. In two cases each in Mozambique and Kenya, we detect areas only on a day other than the designated market day. In one case in Malawi, the validation data lists two market days (Mondays and Fridays), yet we detect areas only on Tuesdays and Saturdays. Since in all these cases the detected areas overlap directly with locations that appear similar to other listed marketplaces we detect in VHR imagery, we update the validation data to feature the detected days instead of the initially designated ones. Fig. S12 illustrates the detected areas for these instances.

**Figure 4.** Across analyses of the activity data, we need to ensure that the readings – which have no naturally interpretable unit – are comparable over time and correctly weighted when aggregated spatially or temporally. Intuitively, we aim to scale the readings such that within each marketplace and day of operation, their value equals 0 on average on non-market days and 100 on average on market days over a reference period. The chosen reference period may differ by application: for comparisons across markets, for instance, care should be taken that each market's chosen reference period should represent a relatively 'normal' year, e.g. without major conflict.

A complication stems from the seasonal distribution not just of market activity, but also the frequency of readings. Images (and thus, activity observations) tend to be more plentiful in seasons with relatively little cloud cover and sparser during the rainy season. A simple average across all readings would overweight the less cloudy seasons and would fail to be comparable across locations with different seasons. To address this, we begin by defining the set of market day observations for market $m$ as $\mathcal{A}_m = \{act_{mt}^{raw} : DoW(t) \in \mathcal{D}_m\}$ with $\mathcal{D}_m$ the set of local market days (e.g. Mondays) and the complimentary set of non-market day observations $\mathcal{A}_m^c$. We then construct an interval mean of the non-market day observations $act_{m,0}^{ref} = \frac{1}{|\mathcal{T}_m^{c,10-90}|} \sum_{t \in \mathcal{T}_m^{c,10-90}} act_{mt}^{raw}$ with $\mathcal{T}_m^{c,10-90} = \{t \in \mathcal{D}_m^c : P_{10}(\mathcal{A}_m^c) \leq act_{mt}^{raw} \leq P_{90}(\mathcal{A}_m^c)\}$, and subtract this from all raw measures, thereby centering the non-market day observations around zero: $\mathrm{act}_{mt}^{\mathrm{demean}} = \mathrm{act}_{mt}^{\mathrm{raw}} - \mathrm{act}_{m,0}^{\mathrm{ref}}$.

We then turn to the market day measures: here, we use the instances of market days throughout a given year $y$ where we do have readings to predict readings on all market days throughout that year. We first define the set of eligible observations as $\mathcal{A}_m^y = \left\{\mathrm{act}_{mt}^{\mathrm{demean}} : t \in \mathcal{D}_m, \mathrm{Year}(t) = y\right\}$ and then fit a penalised univariate smoothing spline with an empirically chosen smoothing parameter:

$$\hat{f}^y = \arg\min_{f \in \mathcal{H}^2} \left\{ \sum_{t \in \mathcal{A}^y} \left( act_{mt}^{demean} - f(t) \right)^2 + \lambda \int \left( f''(t) \right)^2 \, dt \right\}$$

subject to $\sum_{t \in \mathcal{A}^y} \left( act_{mt}^{demean} - \hat{f}^y \right)^2 \leq \frac{|\mathcal{A}_m^y|}{1.5} \, \mathrm{Var}(\mathcal{A}^y)$.

This yields a predicted activity series across market days within a year $act_{m\tau}^{interp,y} = \hat{f}^y(\tau), \qquad \tau \in \{\text{calendar days } \tau : \mathrm{Year}(\tau) = y\}$. We average this series to obtain a year-by-location-specific reference mean on market days $act_{m,1}^{ref,y} = \frac{1}{365} \sum_{\tau=1}^{365} act_{m\tau}^{interp,y}$ and use this to scale all activity measures, across all years and independent of market day status, such that 0 corresponds to the average value on non-market days, and 100 corresponds to typical levels of activity during a specific reference year:

$$act_{mt}^{index,y} = 100 * \frac{act_{mt}^{raw} - act_{m,0}^{ref}}{act_{m,1}^{ref,y}}.$$

Figs. 4A-C show examples of these data indexed to 2018.

**Figure 5.** Fig. 5A shows the time series of the activity measure for three regions of Ethiopia between 2018 and 2024. When considering a time series spanning the shift in sensor configurations between 2020 and 2021, we found that despite our previous harmonisation procedure, activity readings differ between satellite generations. These differences appear idiosyncratic by region, likely reflecting how the different sensors capture the different colours across marketplaces. We therefore aim to harmonize the two series - intuitively, we want the activity series from the two sensors to show the same patterns. At a market level, however, individual readings are too temporally sparse and noisy to perform a direct harmonization. Instead, we derive equal-population subdivisions within regions (Ethiopia's admin-1 level) of up to 2.5 million people and match the means and standard deviations of the two activity series within the subdivisions across the months where each series makes up at least 20% of the total available imagery (mostly between November 2020 and May 2021). Relative to a harmonization at the admin-1 level, this procedure has the advantage of allowing the mismatch between the two activity

series to differ across space. We note that for applications using only values pre-2020 or post-2021, such harmonization will not be necessary.

**Supplementary Figures**

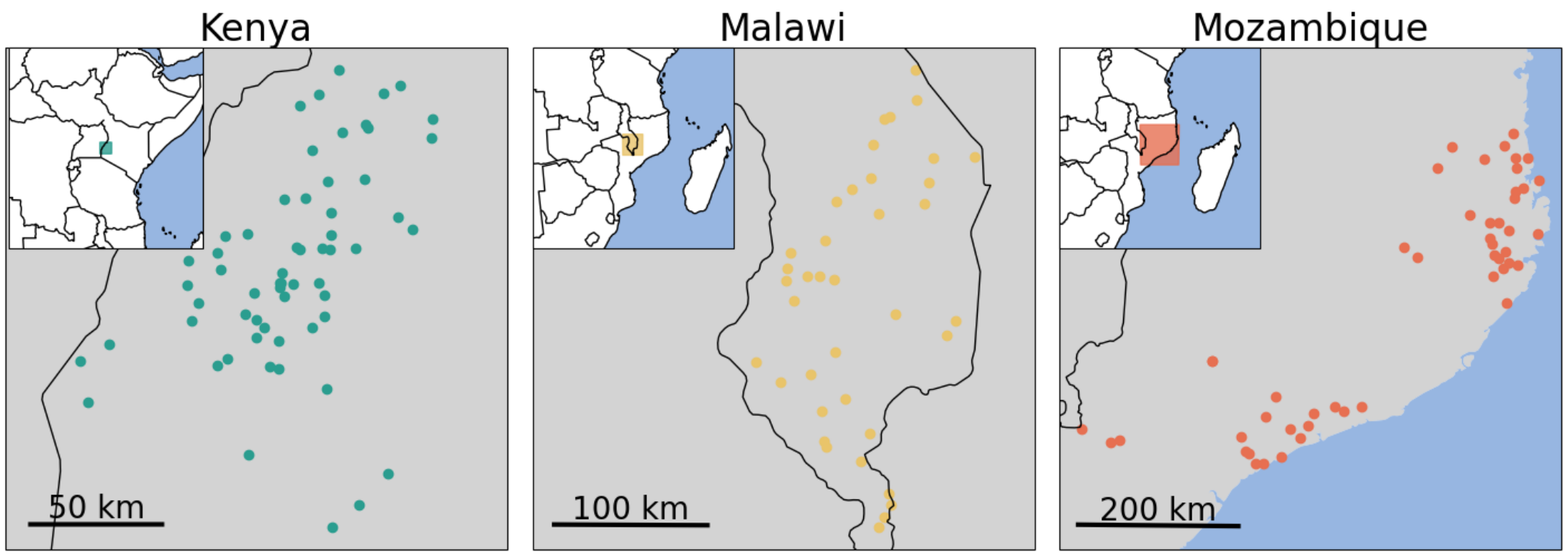

***Fig. S1: Maps of validation datasets for market detection and activity tracking across three countries.*** *Dots indicate locations of periodic marketplaces as recorded in reference datasets.*

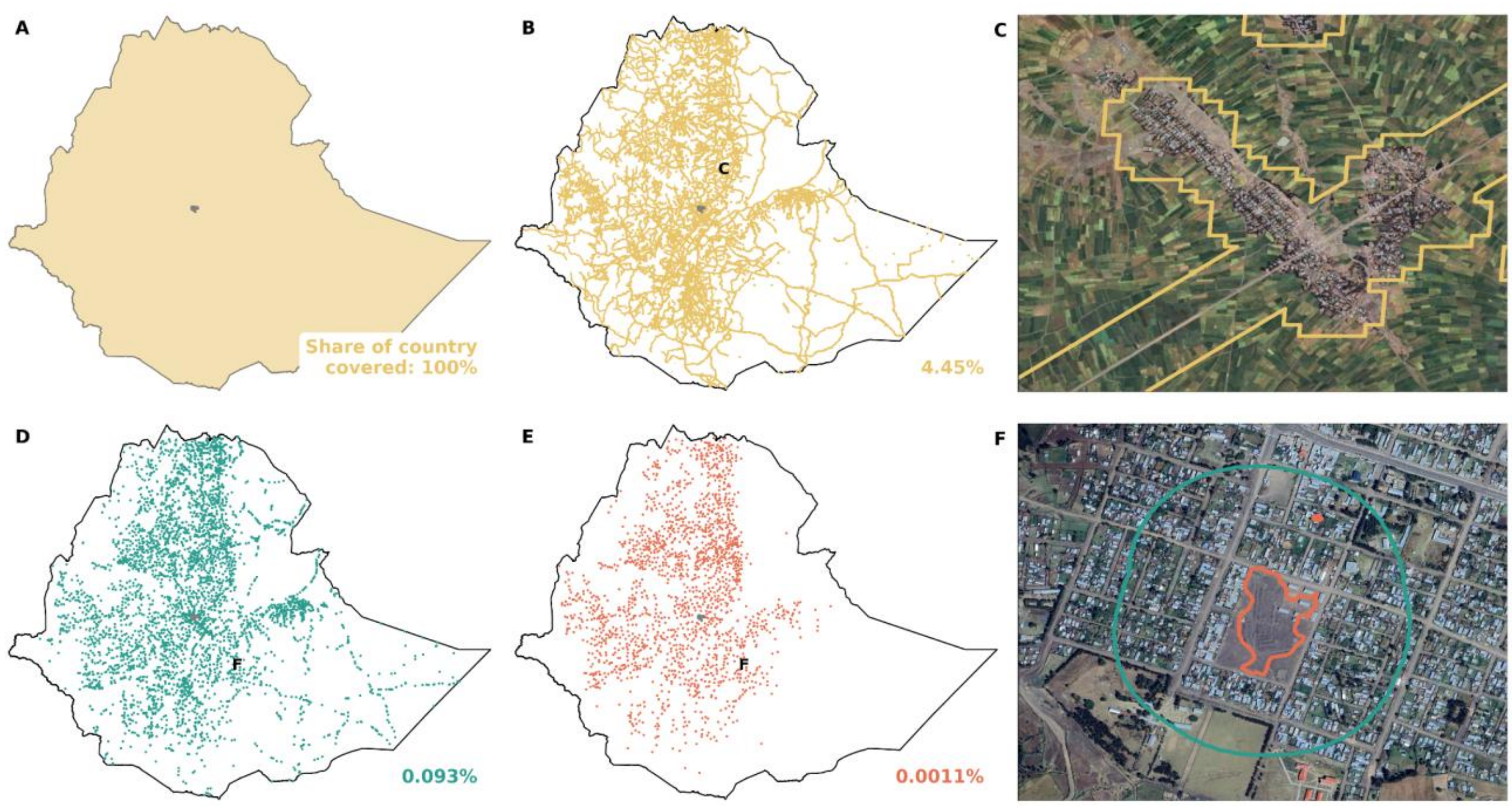

***Fig. S2: Illustration of workflow from candidate location generation to market detection.*** *We use the country-wide GHSL built-up area raster[23] (A) to screen for locations with sufficiently dense settlements. Combined with areas within 300 meters of primary, secondary and tertiary roads in the OpenStreetMap (B), this identifies areas that may reasonably be expected to hold a marketplace. We then screen all these areas (with an example in (C)) for periodic appearance changes using Sentinel-2 satellite imagery, isolating candidate locations at 250 meter radius as shown in green in (F). The 4,400 locations with the highest Sentinel-2 signal are shown in (D),*

*and the subset where we confirmed a marketplace in (E). The bold markers C and F in panels B, D and E indicate the locations of the examples shown in C and F.*

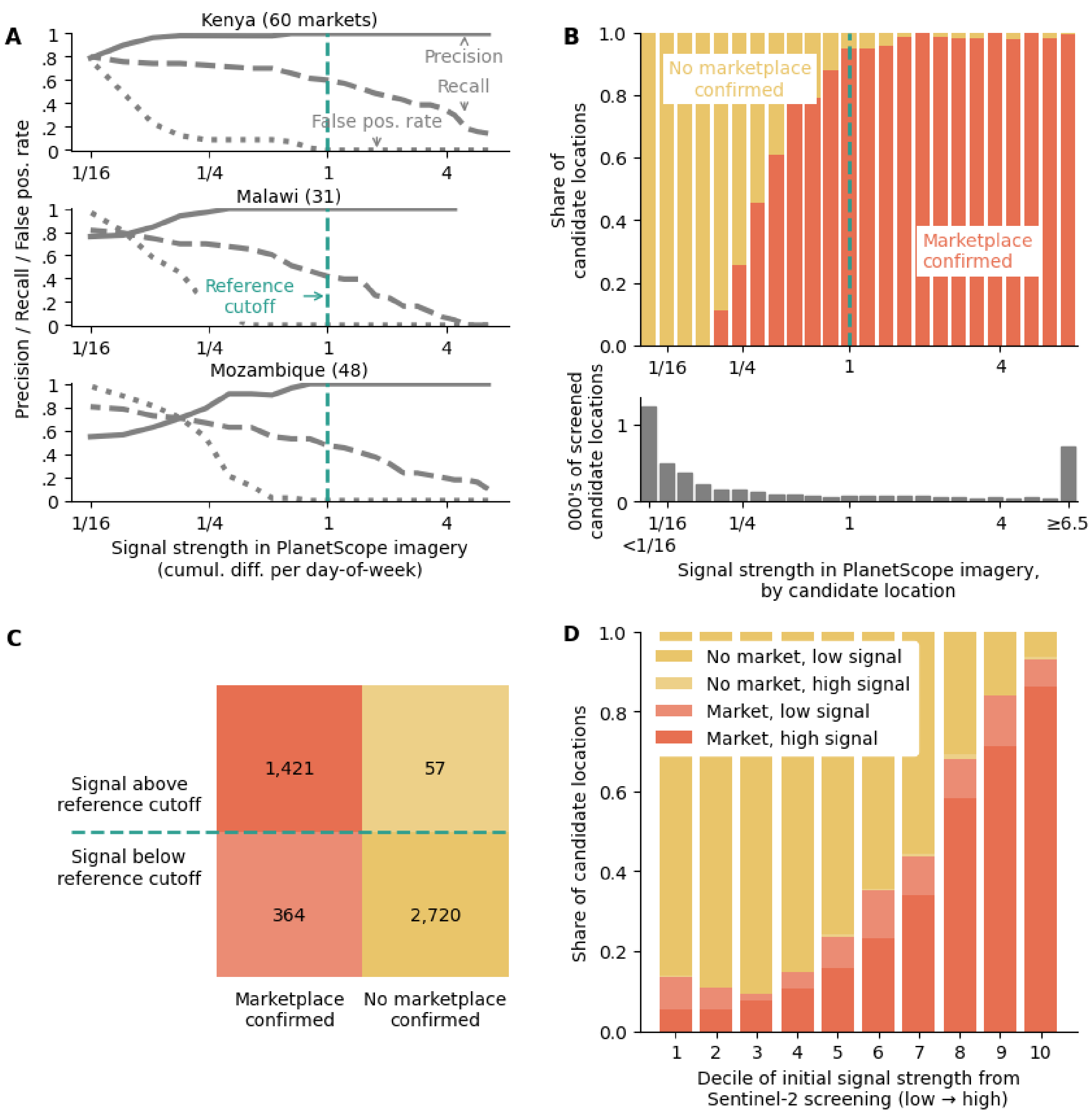

***Fig. S3: Summary of screening procedure for marketplaces in Ethiopia.*** *(A) Repetition of Fig. 2 with accuracy-maximizing threshold for reference. (B) Candidate locations screened with PlanetScope imagery (corresponding to green dots in Fig. S2D), shaded by whether a marketplace was confirmed from visual inspection, or not and sorted by signal strength. The grey bars indicate the distribution of signal strength on the same scale as in (A). (C) Confusion matrix*

*based on accuracy-maximizing threshold. (D) Share of candidate locations returning a marketplace, by decile of Sentinel-2 screening signal.*

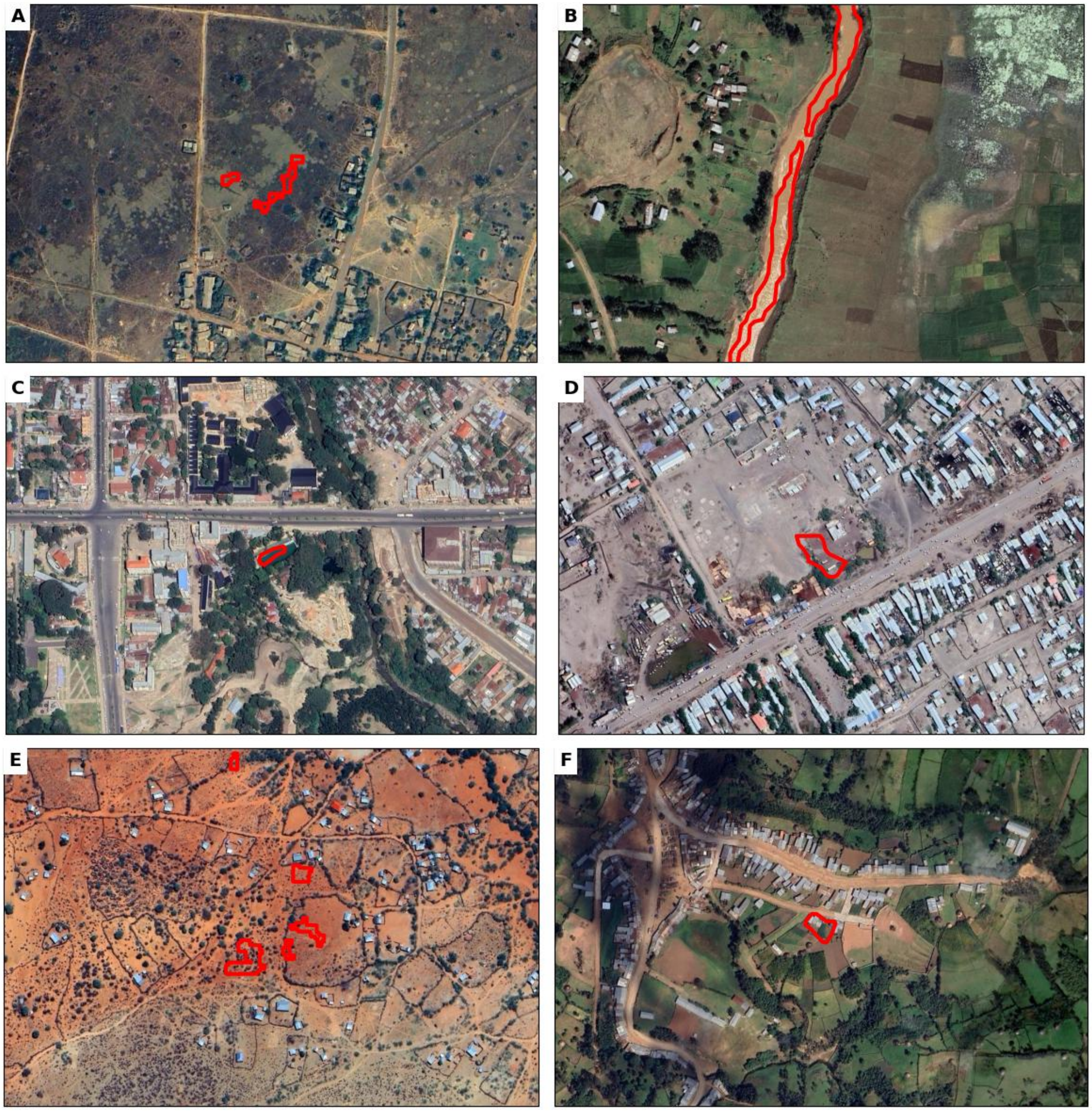

***Fig. S4: Examples of typical detected shapes (in red) not representing marketplaces.*** *The visualizations exemplify the information we use to assess detected shapes as not representing marketplaces, based on their location within the surrounding settlements, as well as their shapes (see online replication package for further examples).* ***(A)*** *6.81°N, 38.61°E. As the identified shapes lie away from public areas and display erratic shapes, we interpret them as reflecting noise.* ***(B)*** *8.33°N, 33.96°E. The detected area coincides with a river, whose changing appearance*

*in the imagery induces substantial noise.* ***(C)*** *5.34°N, 41.88°E. The detected area does not appear to lie in a public area, nor around a covered market building.* ***(D)*** *8.98°N, 37.86°E. Given the location, we interpret the shape as reflecting a periodically-filling parking lot.* ***(E)*** *11.72°N, 40.98°E. Same as (A)* ***(F)*** *11.81°N, 37.64°E. The detected shape coincides with a metal rooftop, which, similar to (B), induces noise.*

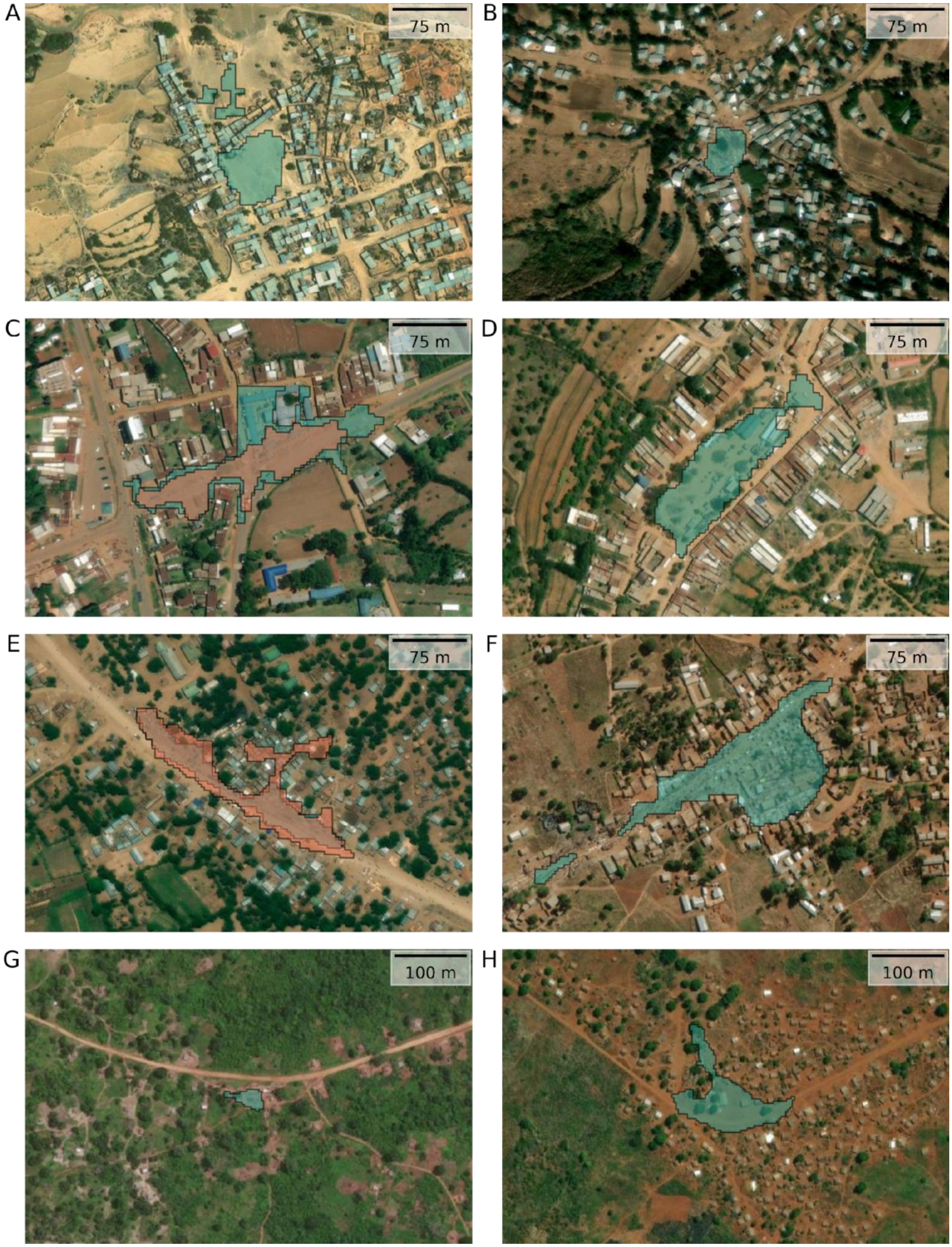

***Fig S5: Examples of detected market areas across four countries. (A)*** *14.35°N, 39.04°E; Ethiopia; Wednesday market* ***(B)*** *11.6°N, 38.44°E; Ethiopia; Saturday* ***(C)*** *0.23°N, 34.52°E; Kenya; Wednesday (green) and Saturday (red)* ***(D)*** *1.62°S, 37.56°E; Kenya; Saturday* ***(E)*** *16.2°S,*

35.02°E; Malawi; Tuesday (green) and Saturday (red) ***(F)*** 15.02°S, 34.76°E; Malawi; Tuesday ***(G)*** 15.05°S, 40.31°E; Mozambique; Saturday ***(H)*** 13.96°S, 40.07°E; Mozambique; Saturday. Basemaps are from ESRI world imagery and show non-market days, except for panels (B) and (F), which are taken on the locations' respective market days. See online replication package for an atlas of detected marketplaces in Ethiopia.*

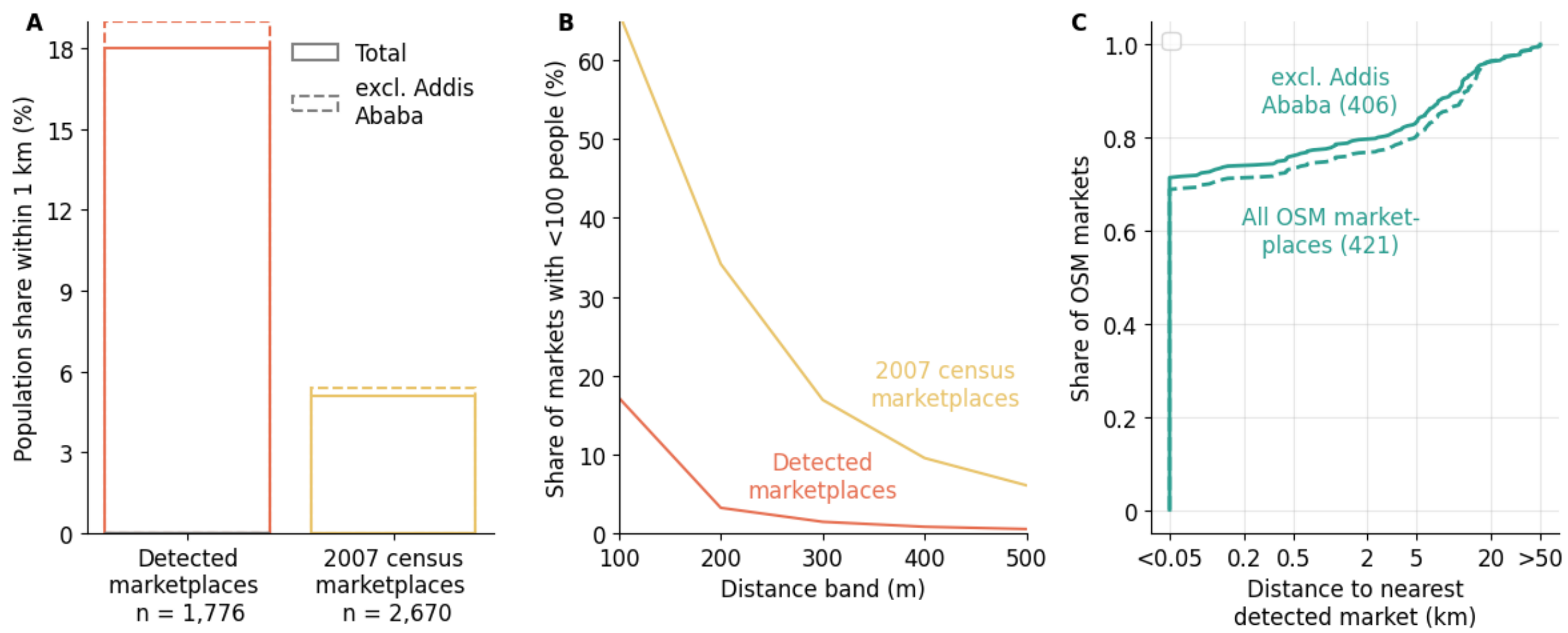

***Fig. S6: Population counts surrounding census marketplaces and distance to OSM-registered marketplaces.*** *(A) Repeating Fig. 3B with marketplaces from the 2007 census. (B) Lines represent the share of marketplaces in the 2007 census and our dataset that have less than 100 people within some distance from their location, according to the GHSL layer[23]. (C) Cumulative shares of locations labelled 'marketplace' in OpenStreetMap by distance to detected marketplaces.*

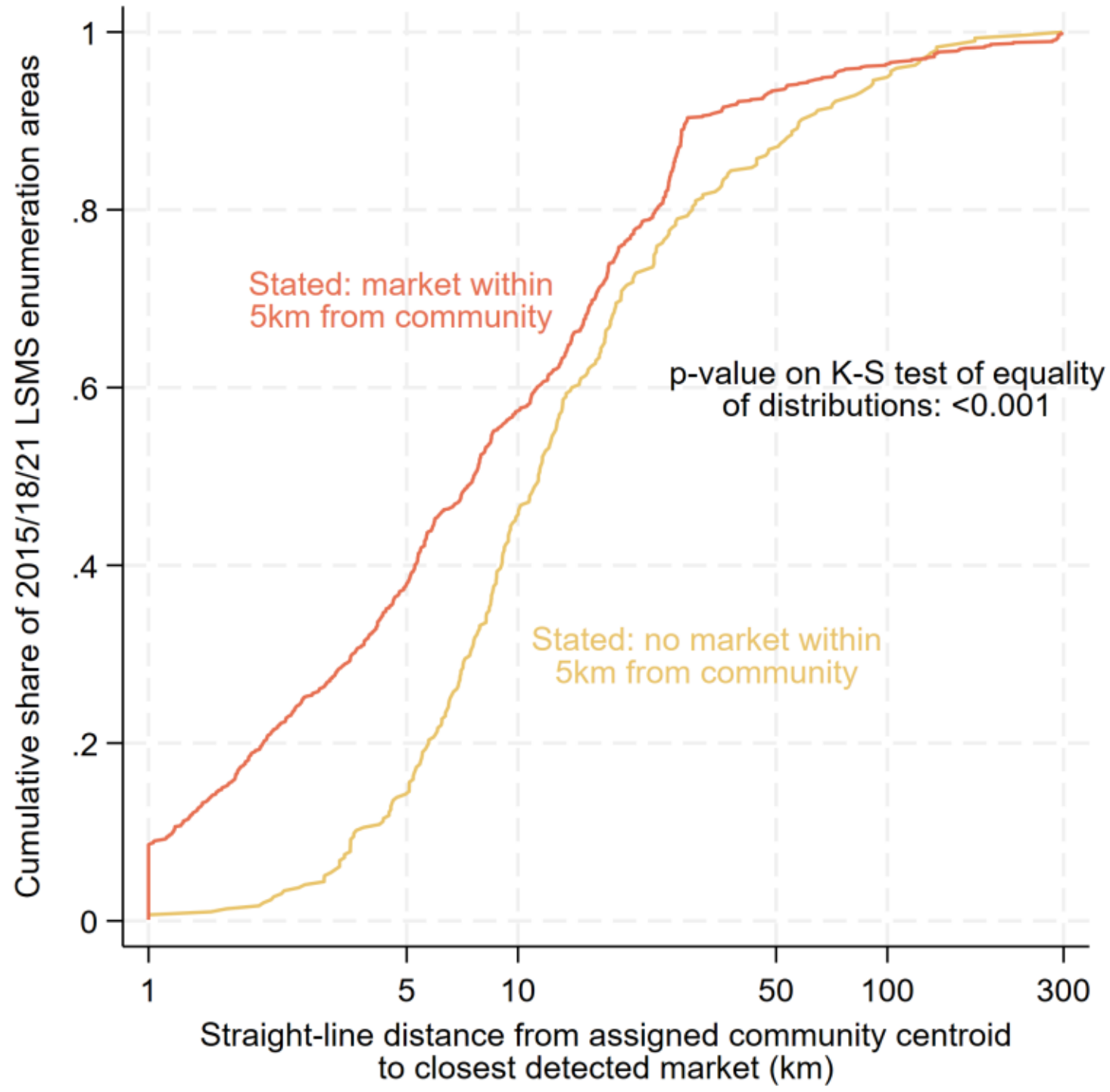

***Fig. S7: Distances between LSMS communities and detected marketplaces, by the stated existence of a 'large weekly market'.*** *Lines represent CDFs of the straight-line distance between the assigned community centroid from the three rounds of LSMS surveys. The assigned centroid is provided from the LSMS surveys, and constructed by averaging the locations of the surveyed households, and subsequently offsetting the mean location by up to 10km. We then construct for each centroid the straight-line distance to the closest detected market in our dataset, and split the sample by communities stating that they have a market either in the community or within 5km, and the remainder.*

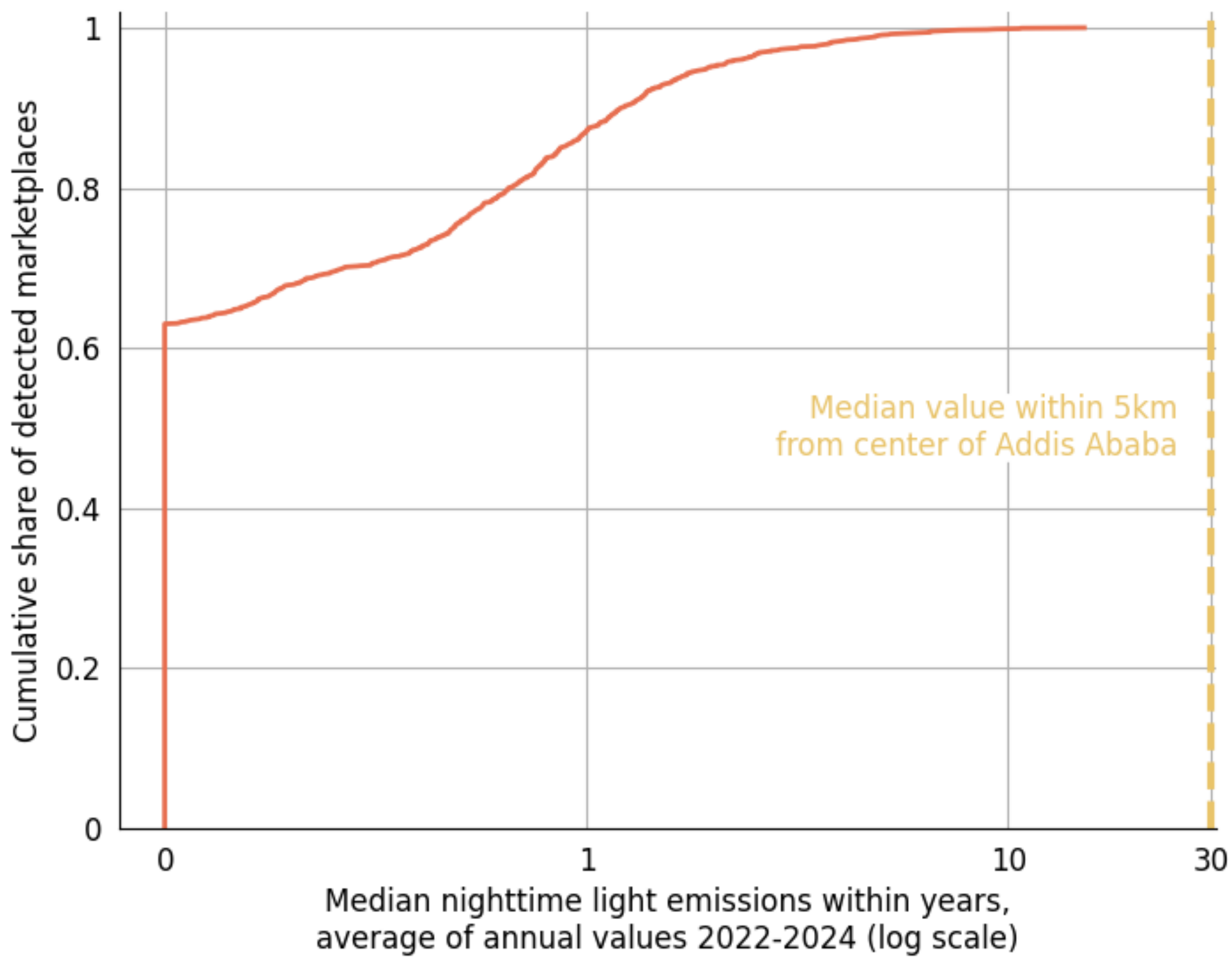

***Fig S8: Cumulative share of marketplaces displaying nighttime light emissions exceeding the indicated level.*** *We spatially match our detected marketplaces to the overlapping gridcell in the annual nighttime light data from* [35]*. These data are, in turn, median composites across monthly VIIRS averages within a year, with various methods for the filtering of background noise. We obtain the data for 2022, 2023 and 2024 and calculate the pixel-level mean across the three rasters. For comparison, the yellow line indicates emissions in a five kilometer buffer around Addis Ababa.*

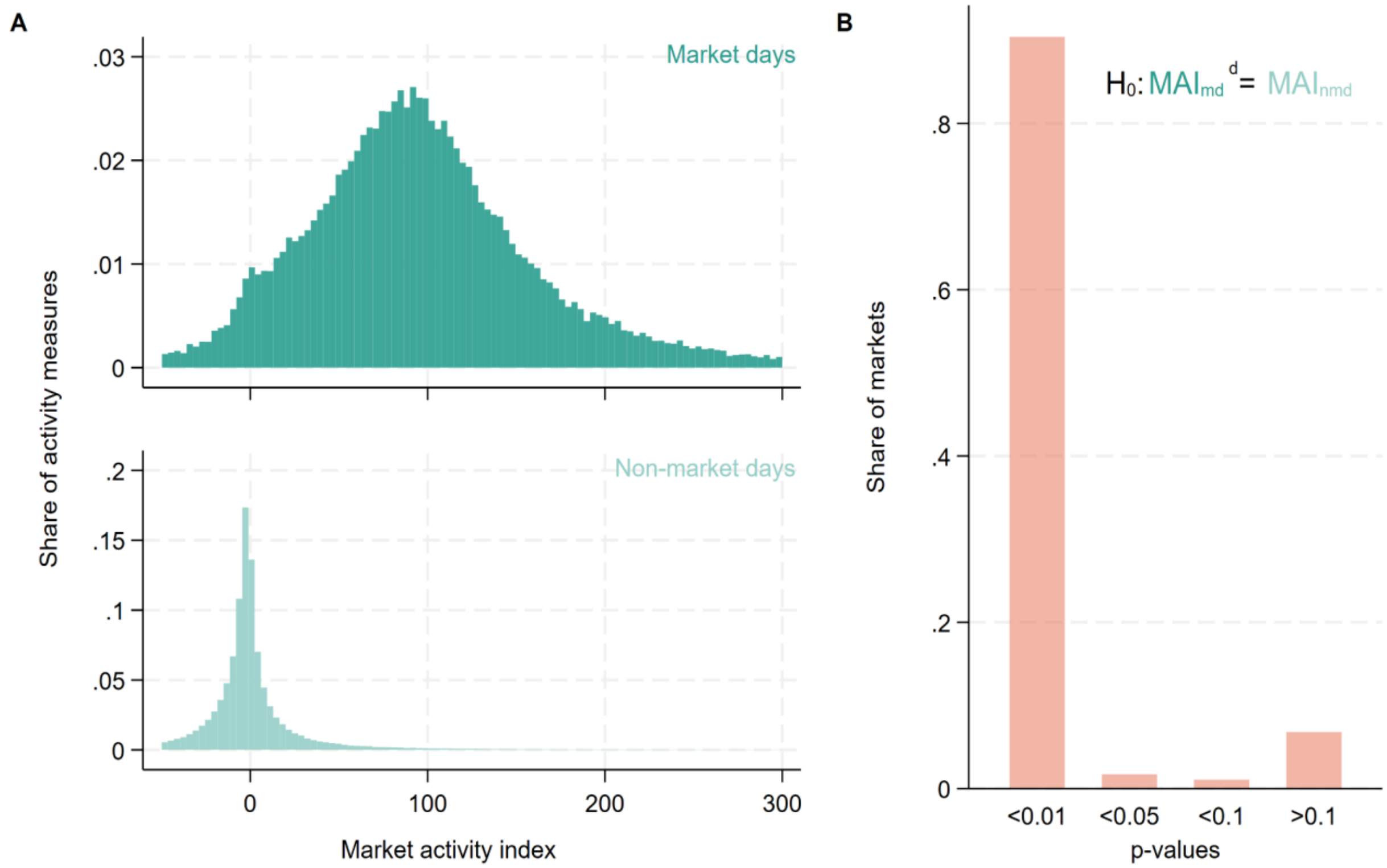

***Fig. S9: Difference in distribution between market day measures and non-market day measures.*** *(A) Histograms of market activity measures in 2018-2019 across detected marketplaces in Ethiopia. The measures are winsorized at (-50, 300). (B) We perform two-sample Mann-Whitney tests for equality of distributions of the activity measures between market days and non-market days within each marketplace. The height of the bars corresponds to the share of marketplaces for which the resulting p-value of the null hypothesis of equal distributions is below / above the respective values.*

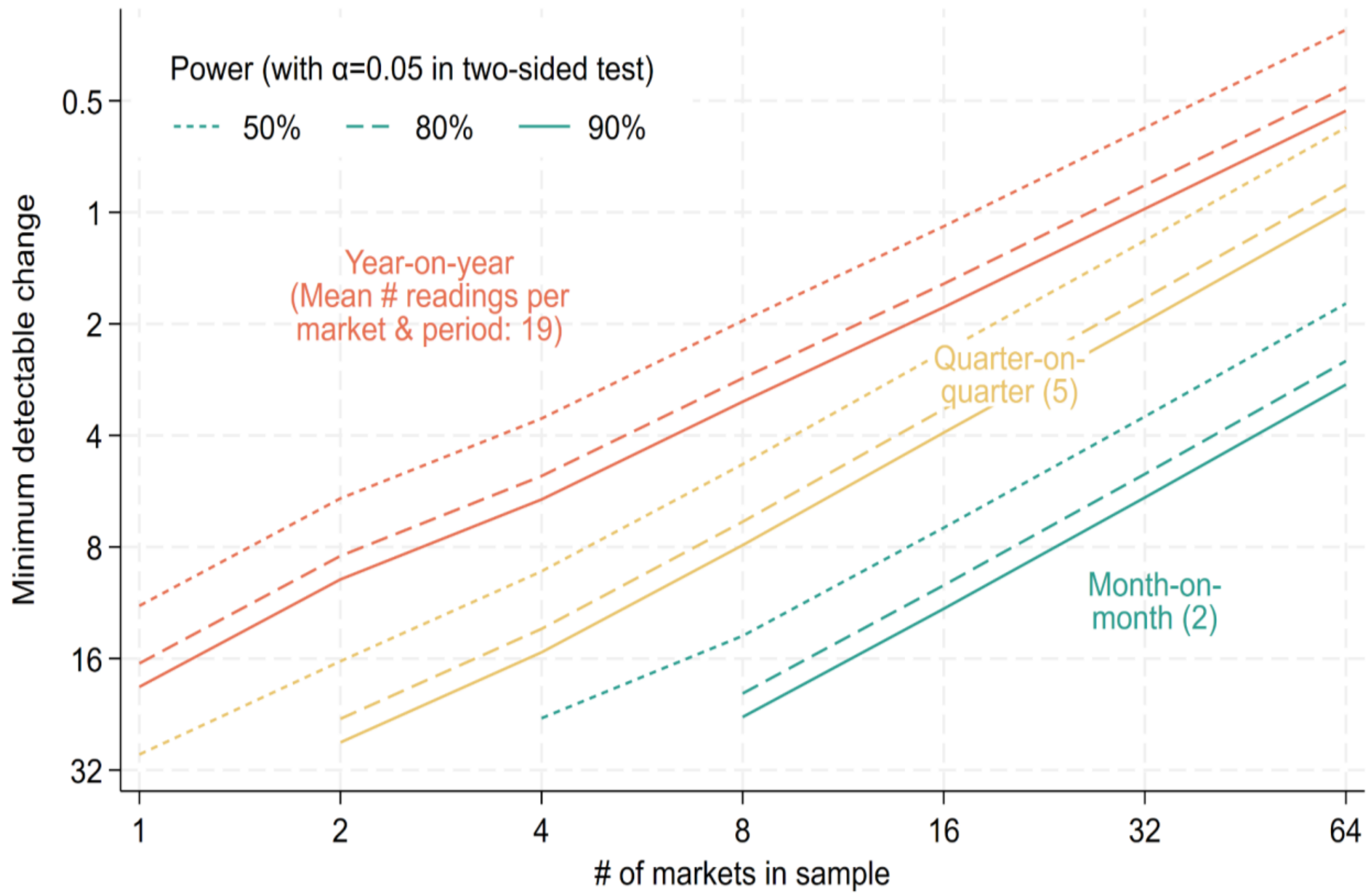

***Fig. S10: Power analysis to detect changes in market activity of various extents (vertical axis) with market samples of varying sizes (horizontal axis), depending on time period of aggregation (colours).*** *Analysis is based on standard deviation of monthly means of market activity from randomly drawn market samples of the specified size in Amhara region for 2018-19.*

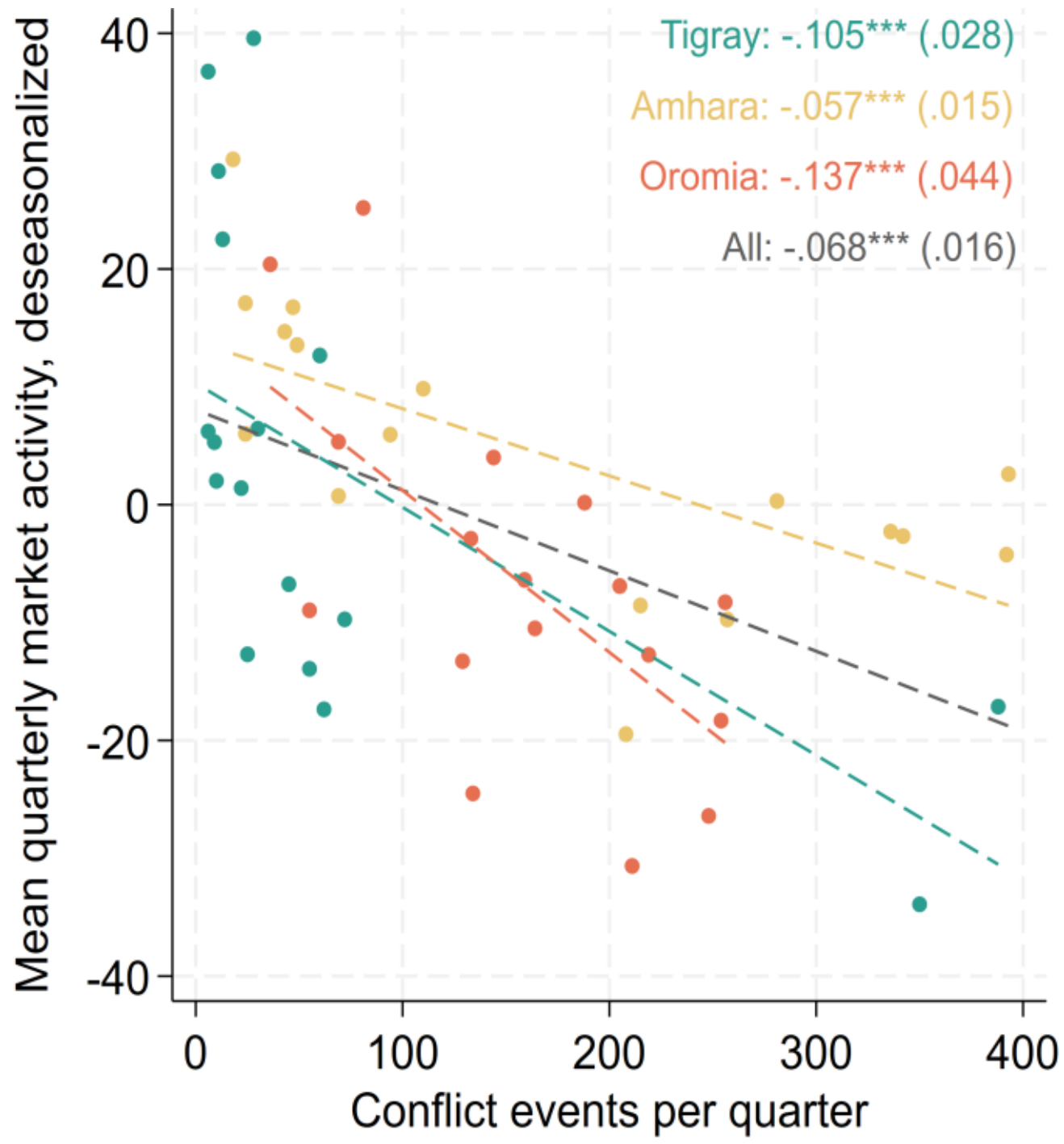

***Fig. S11: Correlations between market activity and conflict, by region.*** *The figure repeats Fig. 5B, with regression lines and estimates shown within and across regions.*

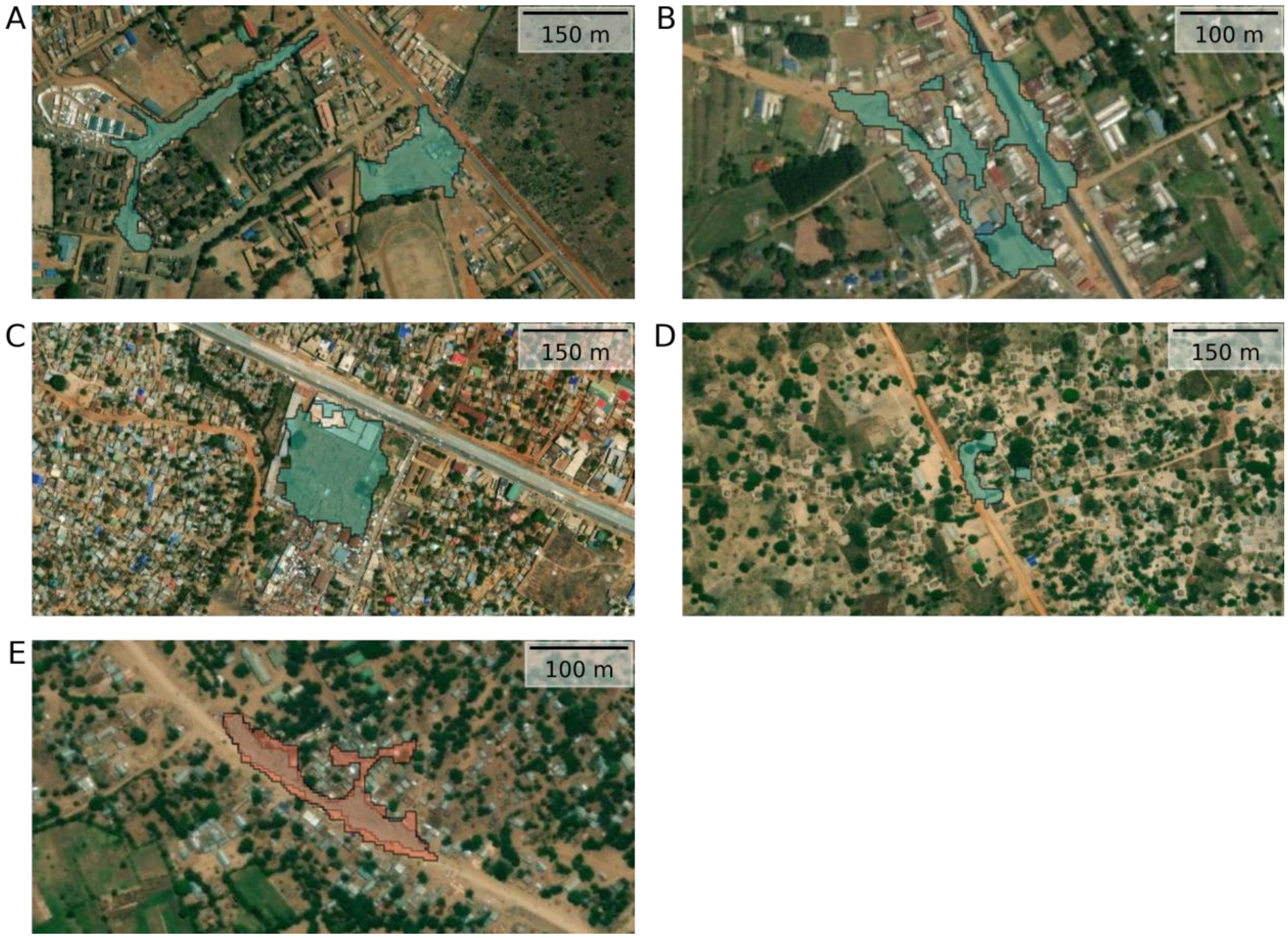

***Fig. S12: Validation markets with reassigned market days.*** *For the shown locations, the validation data listed other market days than the sole ones on which we detect market activity, always in locations that appear highly similar to other marketplaces we detect. We therefore adjusted the validation data in these cases to reflect the high likelihood of falsely listed market days.* ***(A)*** *1.03°N, 35.00°E; Kenya; Tuesday instead of Friday* ***(B)*** *0.49°N, 34.84°E; Kenya; Monday instead of Friday* ***(C)*** *15.13°S, 39.28°E; Malawi; Sunday instead of Tuesday* ***(D)*** *16.97°S, 38.05°E; Malawi; Tuesday instead of Monday* ***(E)*** *16.2°S, 35.02°E; Malawi; Tuesday and Saturday instead of Monday and Friday. Basemaps are from ESRI world imagery and show non-market days.*

## Supplementary Tables

| Detected day: | | any | | Fri, Sat or Sun | |
|---|---|---|---|---|---|
| Number and share of detected areas within … from religious building | 50m | 1 | .06% | 0 | 0% |
| | 100m | 5 | 0.3% | 2 | .3% |
| | 250m | 60 | 3.3% | 22 | 2.7% |
| Total number of detected locations | | 1,776 | | 814 | |
| Distance to closest recorded religious building (km) | Mean | | 16.1 | | |
| | Median | | 12.1 | | |

***Tab. S1: Number and share of detected markets in vicinity of known religious buildings.*** *We identify churches and mosques from OpenStreetMap's place-of-worship layer. The distance is calculated as the straight line between the detected market and the closest religious centre in the area. Friday, Saturday and Sunday are chosen as days when religious services may be held. We cannot reliably identify the denomination of each religious building and their associated days of worship.*

| | | By region | | |
|---|---|---|---|---|
| | (1) | (2) | (3) | (4) |
| | All markets | Tigray | Amhara | Oromia |
| Lean season reduction | -17.772*** | -20.767*** | -16.301*** | -17.911*** |
| | (1.650) | (2.550) | (2.606) | (2.198) |
| Constant | 94.326*** | 95.718*** | 95.829*** | 93.132*** |
| | (1.219) | (1.652) | (1.230) | (2.356) |
| Observations | 48,010 | 5,506 | 19,002 | 17,638 |
| R-squared | 0.152 | 0.112 | 0.122 | 0.182 |
| Market-FE | yes | yes | yes | yes |

***Tab. S2: Reductions in market activity during lean season, compared to normal levels.*** *We separately identify for each administrative zone the three least busy months-of-year in terms of median readings across markets. Similarly, we identify the three busiest months-of-year, and exclude those from the analysis. The regression then compares readings between January 2018*

*and February 2020 in the three least busy months to the remaining six busier months. Standard errors are in parentheses and clustered by month-of-year* ( *** $p<0.01$, ** $p<0.05$, * $p<0.1$).

**Data & code availability**

We access the satellite imagery used in the paper under an academic research user agreement from the provider, which prevents us from making the imagery available directly. We have, however, publicly deposited all code that extracts information from the imagery as well as the derived datasets and the code producing the figures in a GitHub repository: https://github.com/pauldingus/MAI-replication-package.